\documentclass[a4paper,12pt]{article}
\pdfoutput=1
\usepackage{jcappub}
\usepackage[utf8]{inputenc}
\usepackage[T1]{fontenc}
\usepackage{newtxtext,newtxmath}
\usepackage[cal=cm]{mathalpha}
\usepackage{amsmath,amssymb,bm,latexsym}
\usepackage{braket}
\usepackage{graphicx}
\usepackage{subcaption}
\usepackage[export]{adjustbox}
\usepackage{float}
\usepackage{dblfloatfix}
\graphicspath{{Figures/}}
\usepackage{booktabs}
\usepackage{tabularx}
\usepackage{multirow}
\usepackage{multicol}
\usepackage{tikz}
\usetikzlibrary{arrows,arrows.meta,shapes,shapes.geometric,fit}
\usepackage{xcolor}
\usepackage[normalem]{ulem}
\usepackage{soul}
\usepackage{parskip}
\usepackage{placeins}
\usepackage{pgfgantt}
\usepackage{lineno}

\def \HI{{\sc Hi}}

\newcommand{\anoise}{\ensuremath{\alpha\textit{Noise}}}
\newcommand{\mwedge}{\ensuremath{\textit{Wedge}}}
\newcommand{\manalytic}{\ensuremath{\textit{Analytic}}}

\newcommand{\ubase}{\ensuremath{\textit{Unibaseline}}}

\arxivnumber{2605.10928} 
\title{Mitigating residual foregrounds and systematic errors in SKA1-Low AA* EoR observations via Bayesian Gaussian Process Regression}

\author[a,1]{Samit Kumar Pal,\note{Corresponding author.}}\emailAdd{palsamitkumar@gmail.com}
\author[a]{Abhirup Datta,}\emailAdd{abhirup.datta@iiti.ac.in}
\author[c,d]{Aishrila Mazumder, }\emailAdd{aishri0208@gmail.com}
\author[a,b]{and Anshuman Tripathi}\emailAdd{anshumantripathi85@gmail.com}

\affiliation[a]{Department of Astronomy, Astrophysics \& Space Engineering, Indian Institute of Technology Indore, Indore 453552, India}
\affiliation[b]{Department of Physics, Indian Institute of Technology Kharagpur, Kharagpur 721302, India}
\affiliation[c]{Jodrell Bank Centre for Astrophysics, Department of Physics and Astronomy, The University of Manchester, Manchester M13 9PL, UK}
\affiliation[d]{State Key Laboratory of Radio Astronomy and Technology, Shanghai Astronomical Observatory, CAS, 80 Nandan Road, Shanghai 200030, China}

\date{\today}
\abstract{The redshifted 21\,cm line is an emerging tool in observational cosmology that can serve as a direct probe of the intergalactic medium throughout the cosmic timeline. However, the observation of the cosmological 21\,cm signal is extremely challenging in practice, regardless of the scale of interest and redshift, due to the presence of astrophysical foregrounds and residual systematic errors. Machine-learning-based Gaussian process regression\,(ML-GPR) has proven to be the most effective strategy for signal separation in LOFAR and NenuFAR observations to measure the 21\,cm signal power spectrum from the Cosmic Dawn\,(CD) and Epoch of Reionization\,(EoR). In this work, we extend this framework to synthetic CD/EoR SKA1‑Low observations to assess its robustness in mitigating residual foregrounds against instrumental and environmental systematic effects. We use our developed end-to-end realistic simulation pipeline (\textsc{21cmE2E}) for SKA-Low observations. Our 4\,hour tracking simulation includes extragalactic point sources, the AA* telescope configuration, primary beam response, and residual systematics. The modelled errors incorporate antenna-based gain calibration errors, ionospheric phase errors, partial de-mixing of the out-of-field sources, and instrumental noise for 1000\,hours of deep integration time. This work is one of the first demonstrations of the efficacy of the ML-GPR method in a realistic observing scenario. We compare different Bayesian GPR frameworks to assess their ability to suppress residual foreground contamination while minimising signal loss and providing reliable uncertainty estimates. Our results demonstrate that even in a realistic scenario where observations have a combination of residual systematics, ML-GPR can recover the 21\,cm signal robustly within the $2\sigma$ confidence interval across almost the entire range of k-modes sampled.}


\begin{document}
 \maketitle
 \flushbottom
\section{Introduction}
\label{sec:introduction}
The first billion years of cosmic history witnessed the emergence of luminous sources, such as the first stars and galaxies. The copious energetic radiation emitted by these sources transformed the predominantly neutral intergalactic medium\,(IGM) into an ionized state. This transformative period, known as the Epoch of Reionization\,(EoR), remains one of the least understood epochs in the high-redshift Universe. The most promising probe of the EoR era is the redshifted 21\,cm hyperfine transition line from neutral hydrogen, which arises when a ground-state electron undergoes a spin-flip transition. The detection of the cosmological signal of interest originating from the large-scale distribution of primordial neutral hydrogen is a primary objective in observational cosmology. Several current and future radio interferometers have been designed to map the three-dimensional fluctuations of the 21\,cm brightness temperature. These experiments aim to detect and characterise its power spectrum, thereby shedding light on the properties of the first astrophysical sources and the history of reionization. These instruments include the upgraded Giant Metrewave Radio Telescope (uGMRT) \citep{Swarup1991, Gupta2017CSci..113..707G}, the Murchison Widefield Array \citep[MWA;][]{Tingay2013PASA...30....7T, Bowman2013PASA...30...31B}, the Low-Frequency Array \citep[LOFAR;][]{Haarlem2013A&A...556A...2V}, the Hydrogen Epoch of Reionization Array \citep[HERA;][]{Deboer2017}, the New Extension in Nançay Upgrading LOFAR \citep[NenuFAR;][]{Zarka2012sf2a.conf..687Z, Zarka7136773}, and the Square Kilometre Array (SKA) \citep{Koopmans2015}. The SKA, with its unprecedented sensitivity and large collecting area, will significantly advance our understanding of this early cosmic epoch \citep{Koopmans2015, ska_design}.

To date, current radio interferometric 21\,cm experiments have set only increasingly stringent upper limits on the 21\,cm signal power spectrum over a range of reionization redshifts and for a few specific length scales \cite{Mertens2025, Nunhokee2025ApJ...989...57N, Trott2025ApJ...991..211T, Munshi2025a, Hera2026ApJ...998...33A}. One of the major obstacles is the presence of bright astrophysical radio sources, including external galaxies and our Milky Way, as well as spectrally structured systematics. These foregrounds outshine the faint cosmological signal by up to four orders of magnitude. Foregrounds are intrinsically spectrally smooth, whereas the 21\,cm signal is spectrally structured across all scales. In Fourier space, the spectrally smooth foreground emission is confined to a wedge-shaped region and bounded by the horizon line \cite{Datta_2010}. The wedge is a direct consequence of instrumental chromaticity. It results from a mixture of spatial and spectral modes in the cylindrical power spectrum, a phenomenon known as mode-mixing \cite{Morales2012ApJ}. The region of Fourier space expected to be free of foreground contamination is known as the EoR window. Two primary strategies have been applied solely or in a hybrid system to mitigate foreground contamination: a) foreground subtraction, where foregrounds are modelled and removed from the data, and b) the foreground avoidance technique \citep{Datta_2010, Vedantham2012ApJ...745..176V, Thyagarajan2015ApJ...804...14T, Thyagarajan2015ApJ...807L..28T} whereby the wedge-shaped structure is avoided during the extraction of the cosmological 21\,cm signal. However, both strategies are prone to instrumental and environmental systematics that corrupt the spectral coherence of the foregrounds. These include the chromaticity of the instrument, both in the point spread function and in the primary beam \cite{Murphy2024MNRAS.534.2653M, Rath2025MNRAS.541.1125R, OHara2025}, Earth's ionospheric effects, such as refractive shift \cite{Trott2018ApJ, Kariuki2022, Pal2025a}, scintillation, Faraday rotation, low-level radio frequency interference \cite{Wilensky2023}, antenna gain and bandpass stability \cite{Mazumder2022, Mazumder2023}, polarized foreground leakage, incomplete uv coverage \cite{Murray2018ApJ...869...25M} and instrumental noise, which makes it difficult to separate and subtract foregrounds. These residual systematic errors, if not treated carefully, can leak into the EoR measurements, thereby biasing robust interpretations of the observed signal \cite{Datta_2010, Barry2016MNRAS, Kern2019ApJ...884..105K, Mazumder2022, 2025Tripathi}. This work presents a strategy to mitigate these residual systematics via foreground mitigation techniques from the post-calibration data.

The primary goal of foreground subtraction techniques is to model and subtract bright sources from the visibility domain. However, even after this initial subtraction, residual foregrounds and systematic contaminants remain several orders of magnitude brighter than the target signal. Therefore, recovering the underlying cosmological signal of interest within this foreground-dominated wedge requires exploiting the distinct statistical and spectral behaviours of the foregrounds and the 21\,cm signal. Several techniques address this challenge, including blind statistical subtraction, Bayesian frameworks, and unsupervised machine learning architectures \citep{Chapman2015, Mertens2018MNRAS.478.3640M, Kern2021, Mertens2024, Mertens2025, Bianco10.1093/mnras/staf973}. For instance, Mertens et al. \cite{Mertens2018MNRAS.478.3640M} introduced a non-parametric Gaussian Process Regression\,(GPR) method within a Bayesian framework. This framework was later enhanced by replacing the analytic covariance kernel with a variational autoencoder\,(VAE) that learns the shapes of 21\,cm signals across different reionization scenarios \cite{Mertens2024}. This VAE-based GPR has since been applied to LOFAR and NenuFAR observations to tighten the upper limits on the 21\,cm power spectrum \cite{Mertens2025, Munshi2025a} and has emerged as one of the most effective strategies for signal separation \cite{Bonaldi_sdc3a}. In addition, Beohar et al. \citep{Beohar2026}  further demonstrated that the hybrid approach enables the recovery of relevant large scales (i.e., small $k$-modes). However, when gain-calibration errors exceed $0.1\%$, mitigating residual foregrounds with standard GPR models can cause significant signal suppression. Liu et al. \cite{Liu2026_MLGPR} explored different parameterized GPR models for SKA1-Low AA4 observations to identify the most effective modelling strategy and parameter settings. However, their analysis relied on relatively simple simulations that omitted realistic antenna-based and direction-dependent calibration errors, which can substantially affect residual foreground mitigation in actual scenarios. Since these residual systematics introduce spectrally structured features that can mimic or obscure the cosmological signal. Characterizing their impact on Bayesian foreground separation, and validating the ML-GPR technique on realistic mock SKA data that include propagation and instrumental systematics, is therefore essential. In this work, we test the robustness of an ML-enhanced GPR framework on synthetic SKA1-Low AA* EoR observations, accounting for complex astrophysical foregrounds and realistic systematic effects. We compare four Bayesian GPR frameworks for recovering the cosmological 21\,cm signal, and assess which covariance prescription best limits signal loss and residual contamination while yielding reliable uncertainty estimates, using Bayesian model evidence and $z$-score diagnostics.

This paper is organized as follows. In Section~\ref{sec:syn}, we briefly summarize how we set up the SKA simulation framework and describe the sky models, telescope layouts, and error models, including instrumental and environmental effects, used to simulate realistic interferometric observations. In Section~\ref{sec:methodology}, we present our method for generating synthetic maps and power spectrum measurements from the simulation output and discuss how the imaging weighting scheme is used to subtract bright sources. In Section~\ref{sec:foreground_mitigation}, we outline the detailed procedure for subtracting residual foregrounds using ML-GPR. In Section~\ref{sec:results}, we compare the performances of different Bayesian GPR models for extracting the 21\,cm signal from SKA-Low observations. Finally, in Section~\ref{sec:summary}, we summarize the key findings of this study. In this study, we used the $\Lambda$CDM cosmology with the Planck 2018 parameter set \cite{planck2018}: $\Omega_M= 0.31, \Omega_{\Lambda} =0.68, \sigma_{8}=0.811, H_0=67.36$\,\text{km~s}$^{-1}$~Mpc$^{-1}$.

\section{Simulations}\label{sec:syn}
We generate mock radio interferometric observations using the end-to-end \textsc{21cmE2E}\footnote{\url{https://gitlab.com/samit-pal/21cme2e.git}} pipeline \cite{Mazumder2022, Mazumder2023, Pal2025a, Pal2025b}, which is built on the \textsc{OSKAR} simulation software \cite{Dulwich2020zndo...3758491D}. \textsc{OSKAR} uses the Radio Interferometer Measurement Equation\,(RIME) \cite{Hamaker996} to generate full-Stokes visibility data. This pipeline includes realistic sky models, telescope configurations, and the time-varying primary beam response of the SKA to produce realistic SKA observation visibility data and construct synthetic maps from them. In this work, only Stokes-I polarization is simulated to reduce computational cost, thereby neglecting polarization leakage effects on the cosmological signal. The simulated 4-hour tracking observation spans an hour-angle range from $-2\mathrm{h}$ to $+2\mathrm{h}$, centered at right ascension $\alpha = 0\mathrm{h}$ and declination $\delta = -30^{\circ}$. The simulation covers a $138 - 146$ MHz frequency range with a 125 kHz channel width. Each visibility sample is integrated over 10 seconds, resulting in a total of 1440\,time steps for the full observation. The summary of observational parameters is listed in Table~\ref{tab:obs_para}. We additionally inject instrumental noise equivalent to a 1000\,hour deep integration to ensure the noise floor is sufficiently low for the detection of the cosmological 21\,cm signal. A detailed description of the sky and telescope models used in this simulation is provided in the following subsection. This simulation framework allows the inclusion of different types of systematic errors, which we refer to as the error model (see Sec~\ref{error_model}).

\begin{table}
    \centering
    \caption{Overview of the observational parameters used in the simulation framework.}
    \label{tab:obs_para}
    \begin{tabular}{lcccr}
    \hline\hline
    Parameter &&  Value \\
    \hline\hline
    Phase Centre (J2000)     && RA, DEC= 0\,h, $-30\,^{\circ}$\\
    Central frequency      & &142\,MHz (z$\sim 9$)\\
    Bandwidth              & & 8\,MHz\\
    Spectral Resolution         &  & 125\,kHz\\
    Integration time per snapshot && $10$\,sec\\
    Telescope layout   && SKA1-Low AA* \\
    Number of array elements ($N_a$)  && 307\\
    Maximum baseline       & &$\sim 74$\,km\\
    Polarization            & & Stokes I\\
    No. of snapshots        && 1440 \\
    \hline\hline
    \end{tabular}
 \end{table}
\subsection{Sky model}\label{sky-model}
The sky components added in this simulation follow the same prescription 
established by Bonaldi et al. \cite{Bonaldi_sdc3a}. The sky model was divided into two distinct parts: inner and outer sky patches. The inner sky region represents the most beam-sensitive part of the target field of view.
The outer sky model accounts for the remaining hemisphere, covering the full $2\pi$ steradians above the horizon, excluding the in-field patches. This region is critical for modelling the leakage of bright off-axis sources into the station sidelobes. At the lowest frequency channel, the inner sky subtends an angular radius of $4.0^{\circ}$. Because the primary beam response has minimal variation across the simulated bandwidth, we assume a fixed boundary between the inner and outer sky patches that remains constant over the full frequency range for simplicity in this simulation.

We include the cosmological 21\,cm signal and extra-galactic point sources in the inner patch. The extra-galactic sky model was taken from a composite GaLactic and Extra-galactic All-Sky MWA\,(GLEAM)\cite{Wayth2015, Hurley-Walker2017MNRAS.464.1146H} and Long Baseline Epoch of Reionisation Survey\,(LoBES) catalogue \cite{Lynch2021}. The outer patch contains discrete sources from the same GLEAM/LoBES catalogues with $S_{\rm 200\,MHz} > 5$\,Jy. Despite their far-field distance from the pointing centre, these off-axis sources contribute a substantial fraction of the total sky flux intensity budget. All source spectra are modelled as power laws with spectral indices referenced at 200\,MHz, and any sources with invalid spectral indices are discarded. In this simulation, we did not include the Galactic diffuse foreground component. The cosmological 21\,cm signal is not added in the outer sky regions because its contribution to off-axis leakage is negligible compared to that of bright point sources, thereby reducing the computational cost.

To generate the brightness temperature maps of the \HI\ 21\,cm signal at our redshift of interest ($z \sim 9$), we used {\sc 21cmFAST} \citep{Mesinger2011, Murray2020}. This simulation employed a semi-numerical approach based on the excursion set formalism. The simulated physical light-cone covers a comoving sky-plane area of $500\times500$\,$h^{-1}$Mpc$^2$. This physical lightcone (cMpc$^{3}$) is converted from comoving Mpc to angular-frequency coordinates (deg, deg, MHz) for input to the \textsc{21cmE2E} pipeline.

\subsection{Telescope model}\label{telescope}
The low-frequency component of the Square Kilometre Array\,(SKA-Low) is a next-generation radio telescope. It is designed to make tomographic maps of the \HI\ $21$\,cm signal and to measure the 21\,cm power spectrum with high precision \cite{Koopmans2015, Mellema:2015IS}. The SKA1-Low is currently under construction at the Murchison Radio-astronomy Observatory\,(MRO) in Western Australia, and will operate over the  $50\text{--}350$ MHz frequency range. In the full array assembly\footnote{\url{https://www.skao.int/en/science-users/ska-tools/543/ska-subarray-templates-library}}\,(AA4) configuration, the instrument will comprise $131,072$ log-periodic dipole antennas distributed across 512 aperture-array stations. Each station contains 256 dipoles arranged in a Perturbed Vogel pattern \cite{Battaglia2025, Davidson11000012}. This layout breaks regular geometric structure to reduce mutual-coupling resonances while preserving the radial density profile and maintaining high packing efficiency. To reduce computational costs in our simulation, we adopt a single, identical antenna layout for all stations rather than simulating a unique layout for each. As a result, the primary beam pattern of every station is identical in shape (differing only by sky position through pointing and projection), and the array response reduces to the auto-correlation of a single-station beam pattern. This simplification causes station beams to add coherently, enhancing constructive interference and a stronger sidelobe response. To mitigate this effect, we attenuate the flux density of radio sources near the edge of the field of view by a factor of $10^{-3}$ \cite{Bonaldi_sdc3a, Bonaldi_2025arXiv}. This attenuation also assumes suppression of the far-sidelobe response and partially successful source modelling and subtraction in these regions.
\begin{figure}
    \centering
    \includegraphics[width=\linewidth]{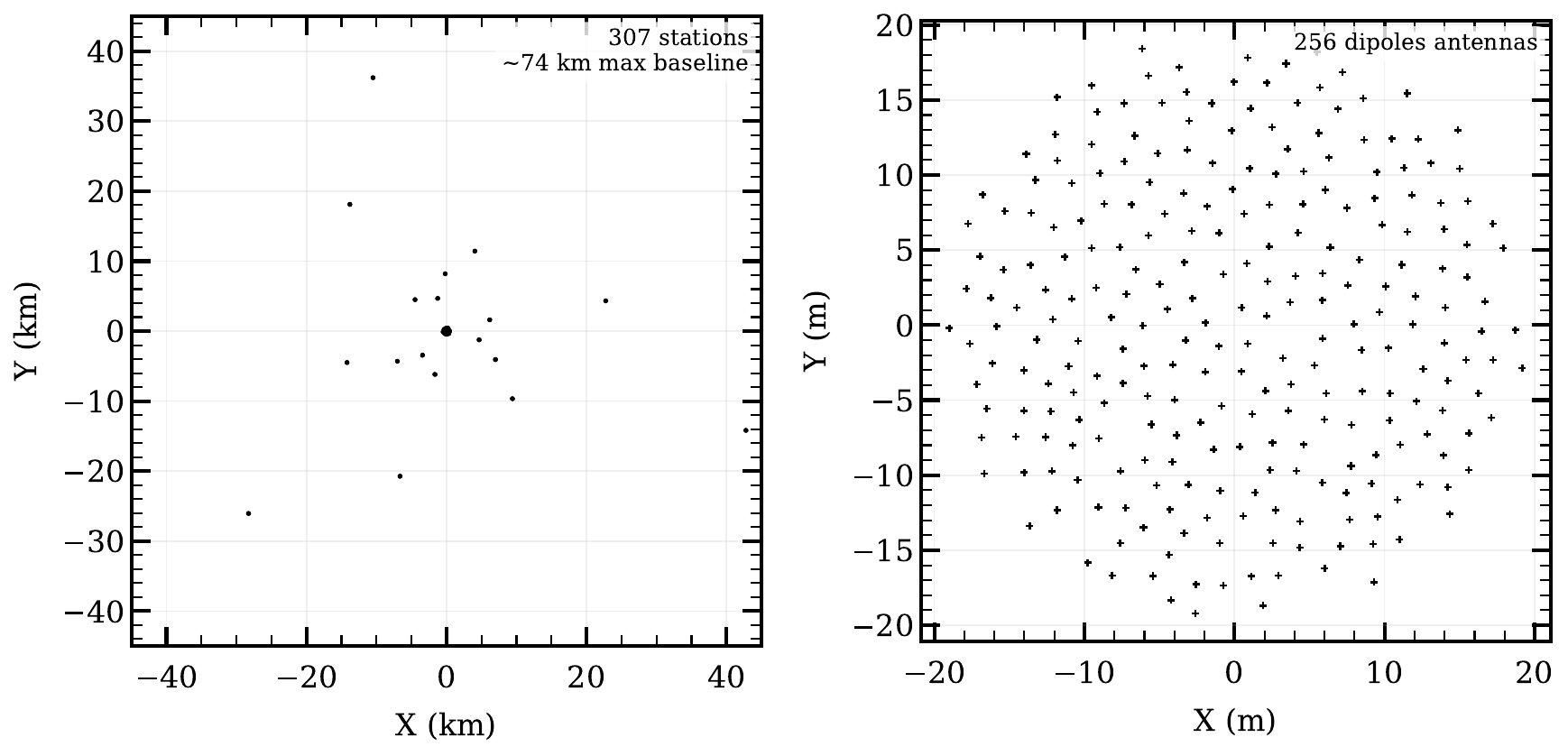}
    \caption{The SKA1-Low AA* array configuration used in this OSKAR simulation. The left panel represents the expected telescope layout for the SKA1-Low AA* configuration with 307 stations. Each station consists of 256 dipole element antennas distributed in a perturbed Vogel pattern. This distribution is used to simulate the station primary-beam response of an SKA1-Low station. The right panel shows the relative positions of the 256 elements within a representative station. In our simulation,  we assume the same element layout distribution for all stations.}
    \label{fig:telescope}
\end{figure}

In this study, we consider the forthcoming SKA1-Low AA$^{*}$ configuration\footnote{\url{https://www.skao.int/sites/default/files/documents/SKAO-TEL-0000818-V2_SKA1_Science_Performance.pdf}}, which comprises 307 stations (at the time of writing) with a maximum baseline of $\sim 74$ km and serves as a subarray of the full AA4 configuration \citep{SKAO_telescope}. In practice, longer baselines are primarily used for point-source modelling, whereas a compact core (baselines within a 2\,km radius of the array centre) is used for EoR power spectrum estimation. Since the 21\,cm signal is strongest on shorter baselines and decreases rapidly with increasing baseline length, most of its power is concentrated on large angular scales\,(equivalently, small $k$-scales). The expected SKA1-Low AA* array configuration used in the simulation is illustrated in Figure~\ref{fig:telescope}.





\subsection{Error model}\label{error_model}
To generate realistic interferometric observation data, this simulation included various instrumental and propagation error models, such as direction-independent and direction-dependent calibration errors, bright off-axis sources far from the phase centre, and instrumental noise. The following subsections describe the different error models used in the analysis.

\subsubsection{Antenna-based calibration error}\label{di_cal_error}
Next-generation radio interferometers are characterized by very wide fields of view, large fractional bandwidths, high sensitivity, and high resolution. In radio interferometric observations, the cross-correlation of voltages between antenna pairs is frequently contaminated by severe time-, frequency-, baseline-, and direction-dependent effects (DDEs). These systematic errors primarily arise from instrumental complexities, such as pointing errors and mutual coupling within the primary beam, as well as atmospheric phenomena, including ionospheric refraction, diffraction, and Faraday rotation \cite{Tasse_2021, Pal2025a, OHara2025}. In the 21\,cm experiment using radio interferometry, these systematic effects limit the extraction of the cosmological 21\,cm signal owing to imperfect calibration. In general, calibration is performed by calculating the Jones matrix under the assumption of a perfectly known sky model. However, limitations such as parameter degeneracies, incomplete sky models, imperfect knowledge of the primary beam, and rapid atmospheric fluctuations result in residual calibration errors. There are two primary components of gain calibration error: direction-independent\,(DI) calibration errors and direction-dependent\,(DD) calibration errors.

The DI calibration of the telescope stations varies with time and frequency. Following the framework outlined by Pal et al. \cite{Pal2025b}, we simulated antenna-based gain errors by adding a noise term to the gain model. This model accounts for specific standard deviations in both the amplitude and phase for each station across the time and frequency domains. The complex antenna gain $g_i$ for antenna i$^{th}$ can be defined as 
\begin{equation}
    g_i (t,\nu) = (1+ \delta a_i(t,\nu) )e^{-i\phi_i (t,\nu)}
\end{equation}
where $\delta a_i$ and $\delta\phi_i$ represent the errors in amplitude and phase, respectively, of the antenna complex gain. The resulting fluctuations were reminiscent of the gain-error patterns observed in the real data \cite{Chokshi2024}. For this simulation, we considered calibration accuracies achievable either from a single observation or by statistically averaging a large number of independent daily observations. Specifically, we simulated a continuous 4\,hour tracking observation but used it to model a 1000\,hour deep integration, which is equivalent in practice to 250\,repetitions of such tracks. The final post-averaging and post-calibration standard deviations utilized in this study were $0.02^\circ$ in phase and $0.02\%$ in amplitude \cite{Bonaldi_sdc3a, Bonaldi_2025arXiv}. Here, the time-domain residuals represent broadband gain calibration errors, whereas the frequency-domain residuals account for fluctuations in the bandpass calibration.

\subsubsection{Ionospheric error}\label{ionosphere_error}
One of the major challenges in lower-frequency radio observations is the accurate calibration of the propagation of ionospheric-induced phase errors in the observed visibilities. The spatio-temporal variation in the ionospheric plasma density introduces additional phase delays on incoming electromagnetic waves that depend on time, frequency, and sky position. These delays arise primarily from transverse fluctuations in the line-of-sight\,(LoS) free-electron column density through the ionosphere, which is generally quantified as the total electron content\,(TEC). Because the TEC evolves across a wide range of spatial and temporal scales, it introduces baseline-dependent effects on interferometric observations. Under the frozen-screen approximation, as the phase screen drifts across the telescope, it causes the differential TEC observed by a stationary array to vary with time. We characterized the statistics of these phase fluctuations using the phase structure function ($D_{\phi}$), which quantifies the phase variance between two ionospheric piercing points as a function of their separation. The power spectrum of electron density fluctuations in the ionosphere generally follows isotropic Kolmogorov statistics \cite{Mevius2016}. For Kolmogorov turbulence, the structure function is commonly approximated as
\begin{equation}
    D_{\phi}(r) \approx \left(\frac{r}{r_d}\right)^{5/3}
\end{equation}
Here, $D_{\phi}$ is expressed in rad$^2$, $r$ is the separation between two pierce points, and $r_d$ is the diffractive length scale, defined as the separation at which $D_{\phi}(r)$ reaches unity.

To model these ionospheric phase errors, we use the \textsc{ARatmospy}\footnote{\url{https://github.com/shrieks/ARatmospy.git}} code to generate a time-varying TEC (phase) screen. We simulated moderately good observing conditions by adopting a diffractive scale of 10\,km. The ionospheric model comprised two slowly evolving layers at plausible elevations that moved in different directions at different speeds. This phase-screen model introduces direction-dependent calibration errors into the simulated visibilities through \textsc{OSKAR}. To simulate residual ionospheric phase errors, we scaled the TEC fluctuations by a factor of $10^{-2}$, representing what remains after a successful direction-dependent calibration \cite{Bonaldi_sdc3a, Bonaldi_2025arXiv}. It is important to note that we did not apply any mitigation techniques to correct the induced direction-dependent calibration errors caused by the ionosphere. Our aim is to inject these residual errors into the simulation. 

\subsubsection{Partial de-mixing of out-of-field sources}\label{demixing_error}
The sidelobes of the SKA1-Low primary beam have significant sensitivity, which can introduce severe imaging artefacts, such as the aliasing of off-axis sources into the main lobe of the SKA1-Low primary beam. Mitigating these effects is a major challenge during calibration, as poor knowledge of the sidelobes of the primary beam, which vary spatially, temporally, and spectrally, complicates accurate subtraction of bright off-axis sources. The process called de-mixing effectively models and subtracts strong far-sidelobe sources during calibration and imaging pipelines. To account for partially successful de-mixing in this simulation, we followed the methodology outlined by Bonaldi et al. \cite{Bonaldi_sdc3a}. We added bright off-axis sources with amplitudes attenuated by $0.1\%$ of their intrinsic flux in the outer sky regions.

\subsubsection{Instrumental noise}
To simulate instrumental noise for the SKA1-Low, we added uncorrelated Gaussian noise, with the noise level scaled to match the anticipated sensitivity of the array in conjunction with the total deep integration time. Following the framework outlined by Pal et al. \cite{Pal2025b}, we simulated instrumental noise that matched the expected system sensitivity for a 1000\,hour integration time. In this SKA simulation, instrumental noise was generated using OSKAR and is added to the visibility data during simulation. 

\section{Methodology}\label{sec:methodology}
In this section, we describe the methodology used to process the synthetic SKA1-Low observations and to estimate the cosmological 21\,cm power spectrum. We first summarize the reconstruction of image cubes from the simulated interferometric visibilities. We then describe the continuum foreground subtraction strategy required to generate residual image cubes for subsequent mitigation analyses. Finally, we present the image-based power-spectrum framework to characterize the statistical fluctuations of the cosmological 21\,cm signal.
\subsection{Imaging}\label{imaging}
The simulated visibilities generated by the 21cmE2E pipeline were stored in the CASA Measurement Set\,(MS) format \cite{McMullin_2007, CASA-Team_2022}. To create synthetic sky maps, these visibilities were gridded, tapered, and Fourier-transformed using the \textsc{WSCLEAN} tool \citep{Offringa2015PASA, Offringa2019A&A...631A..12O}. We sampled the visibilities into 1,000 $w$-layers to properly account for non-coplanar baseline effects and sky curvature \cite{Offringa2019A&A...631A..12O}. The gridding process utilized a Kaiser-Bessel convolution kernel, employing a window function to accurately estimate the contribution of neighbouring samples to each cell on the $uv$-grid. To ensure gridding systematics remained significantly below the expected 21\,cm signal level, we applied a kernel size of 15 and an oversampling factor of 4095 \cite{Offringa2019A&A...631A..12O}. A high-resolution synthetic map was first generated by combining all frequency channels and applying a uniform weighting scheme. This weighting mode maximizes angular resolution and suppresses point spread function\,(PSF) sidelobes, thereby facilitating the identification of small-scale structures and discrete point sources. The specific imaging parameters applied in the \textsc{WSCLEAN} are detailed in Table~\ref{tab:wsclean_parameters}. For high-resolution imaging, we also utilized the following deconvolution parameters: \texttt{niter = $10^{6}$, mgain = 0.8, automask = 4, and auto-threshold = 1.}
\begin{table}
    \centering
    \begin{tabular}{ccc}
        \hline\hline
         Configuration & Point source subtraction  & GPR removal \\
         \hline\hline
         weight & uniform  & natural \\
         scale &  2048  & 512 \\
         resolution & 16\,arcsec & 32\,arcsec \\
         taper-gaussian & 60  & 60 \\
         taper-edge & 100 & 100\\
         padding & 2 & 2 \\
         wstack-nwlayers & 1000 & 1000 \\
         wstack-oversampling & 4095 & 4095 \\
         wstack-grid-mode & kb & kb \\
         wstack-kernel-size & 15 & 15 \\
         \hline         
    \end{tabular}
    \caption{WSCLEAN parameters used to make a synthetic map.}
    \label{tab:wsclean_parameters}
\end{table}

\subsection{Continuum emission subtraction}\label{compact_source_sub}
The processed images were subsequently corrected using the frequency-dependent primary beam model of SKA1-Low. This primary beam pattern was simulated with \textsc{OSKAR} via the \textsc{oskar\_sim\_beam\_pattern} tool, utilizing the identical telescope configuration employed in the visibility simulations. The beam was evaluated over the full observing band and averaged over 10\,seconds to match the integration time of the observation. The image size and angular resolution were kept consistent with the uniform-weighted image described previously (see Sec~\ref{imaging}).
Using \textsc{PyBDSF}\footnote{\url{https://pybdsf.readthedocs.io/en/latest/}} \cite{PyBDSF2015ascl.soft02007M}, the primary-beam-corrected image was analyzed to compile a source catalogue of the field. This catalogue served as the input for the \textsc{21cmE2E}-pipeline to generate model visibilities, which were then subtracted from the simulated SKA1-Low AA* visibilities using the \textsc{UVSUB} task in \textsc{CASA}. This procedure effectively subtracted the continuum emission from the dataset, yielding residual visibilities with substantially reduced foreground power. Following this continuum subtraction, a dirty image cube was generated for each frequency channel using \textsc{WSCLEAN} tool \cite{Offringa2019A&A...631A..12O}. These cubes were produced using the parameters detailed in Table~\ref{tab:wsclean_parameters}, but applied a natural weighting scheme to maximize sensitivity to the faint 21\,cm signal. To mitigate noise bias during the final power spectrum estimation, we generated two independent dirty image cubes using even- and odd-timestamped data. 

\subsection{Power spectrum estimation}\label{ps_est}
The primary objective of a CD/EoR telescope is to detect the power spectrum\,(PS) of the cosmological \HI\ 21\,cm signal, which quantifies its underlying statistical fluctuations. While the PS can be estimated directly from measured visibilities using delay-spectrum methods \cite{TRott2020, Kern2021, Mazumder2022}, doing so ties each delay bin to a fixed physical baseline, leading to a coupling between spectral and transverse structures \citep{2019MNRAS.483.2207M, Mertens2020}. To minimize this mode coupling and enhance the separation of the cosmological signal from foregrounds, we instead employ an image-based PS estimator \cite{Mertens2020}. This approach applies a Fourier transform along the line of sight at fixed angular scales on gridded image cubes, ensuring strict orthogonality between transverse ($k_\perp$) and line-of-sight ($k_\parallel$) modes.

\begin{figure}[!htbp]
    \centering
    \includegraphics[width=\linewidth]{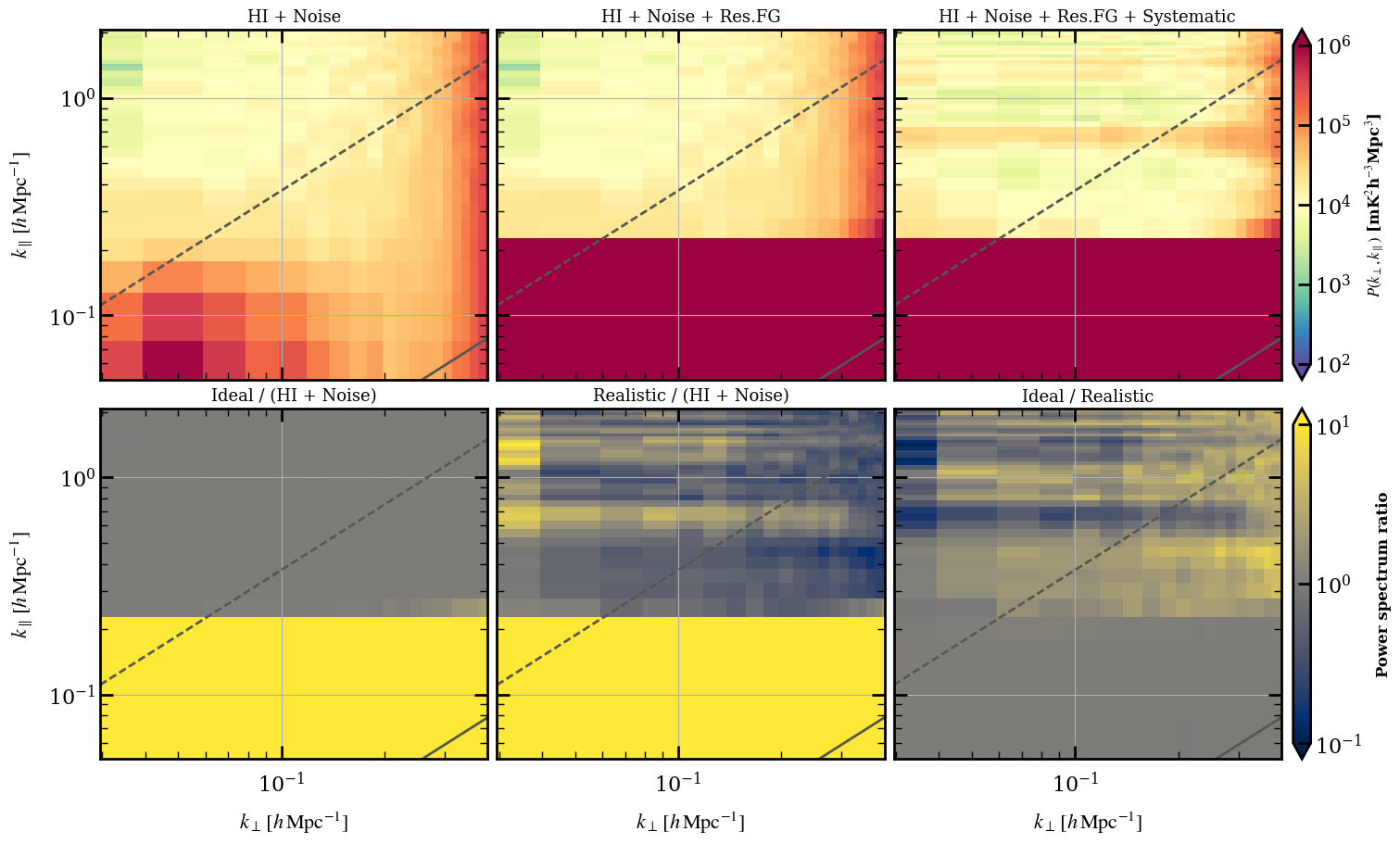}
    \caption{Overview of the cylindrically averaged 2D power spectra $P(k_\perp,k_\parallel)$ of simulated sky components at 138 -- 146 MHz, estimated from the central $3^{\circ}$ field of view. From top left to right: the cosmological 21\,cm signal plus the 1000-hour SKA1-Low AA* thermal noise limit ($P_{\text{HI}+\text{Noise}}$), residual sky after point-source subtraction under ideal calibration ($P_{\text{Ideal}}$), residual sky after point-source subtraction under realistic, imperfect calibration ($P_{\text{Realistic}}$), incorporating antenna-based gain calibration errors, ionospheric phase errors, off-axis sidelobe leakage, and instrumental noise for 1000\,hours of deep integration time. From bottom left to right, compare the ratio between residual sky under perfect calibration to $P_{\text{HI}+\text{Noise}}$, residual sky under imperfect calibration to $P_{\text{HI}+\text{Noise}}$ and residual sky under imperfect calibration to perfect calibration. The bottom-right depicts that residual systematic errors dominate the mock observation over thermal noise. Dashed and solid black lines represent the horizon limit and the boundary corresponding to emission within the primary beam of the telescope, respectively.}
    \label{fig:eor_sky_ps2d}
\end{figure}

For this image-based analysis, we utilize the \textsc{ps\_eor}\footnote{\url{https://gitlab.com/flomertens/ps_eor}} Python package. This tool converts reconstructed image cubes into gridded visibilities and estimates their power spectra. The power spectrum, as a function of wavenumber $k$, is defined as:
\begin{align}
    P(\mathbf{k}) &= \mathbb{V}_{\mathrm{c}} \, \left| \tilde{T}(\mathbf{k}) \right|^2, \\
    \tilde{T}(\mathbf{k}) &= \frac{1}{N_l \, N_m \, N_\nu} 
    \sum_{\mathbf{r}} T(\mathbf{r}) \, e^{-2\pi i \, \mathbf{k} \cdot \mathbf{r}}
    \label{eq:dft}
\end{align}

\begin{equation}
    \mathbb{V}_{\mathrm{c}} = 
    \frac{N_l \, N_m \, N_\nu \; d\ell \, dm \, d\nu \; D_M^2(z) \, \Delta D}
    {A_{\mathrm{eff}} \, B_{\mathrm{eff}}}
    \label{eq:cosmic_volume}
\end{equation}

\begin{align}
    A_{\mathrm{eff}} &= \left\langle A_{\mathrm{pb}}^2(l,m) \, A_w^2(l,m) \right\rangle \\
    B_{\mathrm{eff}} &= \left\langle B_w^2(\nu) \right\rangle
\end{align}
where $\tilde{T}(\mathbf{k})$ is the Fourier-transformed brightness temperature field $T(\mathbf{r})$, and $\mathbb{V}_{\mathrm{c}}$ is the comoving cosmological volume. This volume is normalized by the instrument's primary beam $A_{\mathrm{pb}}(l,m)$, the spatial tapering function $A_w(l,m)$, and the spectral tapering function $B_w(\nu)$. The spatial and frequency dimensions are denoted by $N_l$, $N_m$, and $N_\nu$, respectively. The transverse comoving scales are given by $L_i = N_i \, \Delta i \, D_M(z)$, where $i \in \{l, m\}$, while the line-of-sight comoving depth is $L_\nu = N_\nu \, \Delta \nu \, D_C(z)$ \citep{Morales2004, Mertens2020}. The one-to-one mapping between the cosmological space ($k_x, k_y, k_z$) and instrumental space ($u,v, \eta$) is given by 
\begin{equation}
    k_x = \frac{2\pi u}{D_M(z)}, \quad 
    k_y = \frac{2\pi v}{D_M(z)}, \quad 
    k_z = \frac{2\pi H_0 \, \nu_{21} \, E(z)}{c \, (1+z)^2} \, \eta
    \label{eq:k_components}
\end{equation}
where $\eta$ represents the geometric delay (the frequency Fourier dual), $H_0$ and $c$ are the Hubble constant and the speed of light, respectively, $\nu_{21}$ is the rest-frame frequency of the \HI\ hyperfine transition, and $E(z)$ is the evolution function for the Hubble parameter. The cylindrically averaged power spectrum can be computed from the 3D Fourier representation of the signal by averaging over the transverse directions:
\begin{align}
    k_\perp = \sqrt{k_x^2 + k_y^2}, \quad 
    k_\parallel = k_z
\end{align}
Furthermore, assuming isotropy, the power spectrum can be averaged in spherical shells defined by $k = \sqrt{k^2_\perp + k^2_\parallel}$. The spherically averaged dimensionless power spectrum is defined as:
\begin{eqnarray} \label{eq:dimensionless_ps1d}
    \Delta^2(k) = \frac{k^3}{2\pi^2}P(k)
\end{eqnarray}

Figure~\ref{fig:eor_sky_ps2d} shows the cylindrically averaged 2D power spectra $P(k_\perp, k_\parallel)$ of simulated sky components from SKA1-Low AA* observation. This figure quantifies the cumulative impact of systematic coupling between instrumental/environmental effects and residual foregrounds on the EoR signal. The top row shows the power spectra of the cosmological target signal combined with the 1000-hour thermal noise limit ($P_{\text{HI}+\text{Noise}}$) (\texttt{top-left}), the residual sky after continuum point-source subtraction under ideal calibration ($P_{\text{Ideal}}$) (\texttt{top-middle}), and the residual sky under imperfect calibration ($P_{\text{Realistic}}$) (\texttt{top-right}). From bottom left to right, compare the ratio between residual sky under perfect calibration to $P_{\text{HI}+\text{Noise}}$, residual sky under imperfect calibration to $P_{\text{HI}+\text{Noise}}$ and residual sky under imperfect calibration to perfect calibration. Under perfect calibration, intrinsic foreground emission remains confined to the wedge region at low $k_\parallel$, leaving a relatively clean window for EoR detection at $k_\parallel \approx 0.2\,h\,\mathrm{Mpc^{-1}}$ (bottom-left panel of Fig.~\ref{fig:eor_sky_ps2d}). Under imperfect calibration, however, residual systematics from instrumental and environmental effects scatter power out of the wedge and into the EoR window (bottom-middle panel). The comparison in the \texttt{bottom-right} panel demonstrates that the mock SKA-Low observations are dominated by systematic errors rather than thermal noise across the Fourier region of primary interest. In the following section, we detail our strategy for mitigating these residual systematic and foreground contaminations using the ML-GPR framework.

\section{Residual foreground mitigation}\label{sec:foreground_mitigation}
Even after point-source subtraction, residual unmodelled foreground contamination dominates the simulated SKA data, exceeding the thermal noise level by several orders of magnitude. In our simulated SKA1-Low observations, the target field remains contaminated by confusion-noise-limited foreground emission, residual calibration effects (arising from antenna-based gain and ionospheric errors), and unmodelled flux from bright off-axis sources (due to partial demixing) entering through their PSF sidelobes. These residual foregrounds can nonetheless be distinguished from the 21\,cm signal by exploiting their spectral smoothness, compared to the intrinsic spectral fluctuations of the 21\,cm signal along the line of sight. To model and subtract these residual contributions, we apply Gaussian Process Regression\,(GPR) to the gridded visibility cube in $uv$-space. In this section, we summarize the mathematical principles of GPR-based foreground mitigation before detailing the application of four Bayesian GPR models to our simulated observations.
\subsection{Gaussian process regression}\label{sec:gpr}
We used the machine-learning enhanced Gaussian process regression\,(ML-GPR) method \cite{Mertens2024} for residual foregrounds removal from the pre-processed SKA data (Section~\ref{sec:methodology}). This approach has been proven effective in recent radio-interferometric observations, enabling progressively tighter upper limits on the 21\,cm power spectrum \cite{Mertens2025}. In the context of 21\,cm cosmology, GPR models the gridded visibility data $\mathbf{d}$ in $uv$-space as a sum of continuous Gaussian processes along the frequency axis. These processes represent the foregrounds ($\mathbf f_{\rm fg}$), 21\,cm signal ($\mathbf f_{\rm 21}$), and thermal noise ($\mathbf n$). The SKA simulated observed data can thus be described by, 
\begin{equation}
    \mathbf d = \mathbf f_{\rm fg}(\nu) + \mathbf f_{\rm 21}(\nu) +  \mathbf n(\nu)
\end{equation}
We characterized each component using a frequency–frequency covariance kernel that captures its spectral correlations. The combined covariance kernel function of the data (K) is the sum of these individual contributions and expressed as,
\begin{equation}
   \mathbf K = \mathbf K_{\rm fg} + \mathbf K_{\rm 21} +  \mathbf K_n
\end{equation}
where $\mathbf K_{\rm 21}$ and $\mathbf K_{n}$ represent the 21\,cm signal and noise covariance kernels, respectively. The foreground covariance, $\mathbf{K}_{\rm fg}$, is further decomposed into an intrinsic term $K_{\text{int}}$ and a mode-mixing term $K_{\text{mix}}$. The intrinsic component, $K_{\text{int}}$, accounts for the residual foreground power in the primary field of view and has extremely large-frequency coherence scales. The mode-mixing component, $K_{\text{mix}}$ captures the off-axis foreground sources that occupy the rest of the foreground wedge created by the chromatic and non-ideal $uv$-coverage. The mode-mixing effects introduced by the instrument beam populate the `wedge' feature in the cylindrical power spectrum, as illustrated in the left panel of Figure~\ref{fig:eor_sky_ps2d}.

Previous GPR-based studies in the context of 21\,cm cosmology have demonstrated that the \text{Mat\'ern} kernels are well suited for modelling the spectral structure of both intrinsic foregrounds and mode-mixing contaminants \cite{Mertens2018MNRAS.478.3640M, Mertens2020, Kern2021}. The general form of this class of kernels is given by
\begin{equation} \label{eq:matern}
        \mathrm{K_{\text{Mat\'ern}}}(\nu,\nu') = \sigma^2 \frac{2^{1-\eta}}{\Gamma(\eta)} (\sqrt{2\eta}\frac{|\nu-\nu'|}{l})^\eta K_\eta (\sqrt{2\eta}\frac{|\nu-\nu'|}{l}),
\end{equation}
where $\sigma^2$ represents the variance of the signal, $\eta$ controls the spectral smoothness, and $l$ denotes the characteristic length-scale (i.e., the spectral coherence scale). Here, $\Gamma$ and $K_\eta$ correspond to the gamma function and the modified Bessel function of the second kind, respectively. The hyperparameters in Equation~\ref{eq:matern} make the \text{Mat\'ern} kernel family highly flexible, enabling it to capture a wide range of spectral behaviours. In particular, increasing the parameter $\eta$ from $1/2$ to infinity yields progressively smoother functions, corresponding to signals with increasing frequency coherence. For a more detailed discussion on the roles and effects of the other hyperparameters, we refer the reader to \cite{Mertens2018MNRAS.478.3640M, Mertens2020, Kern2021}.

We modelled the covariance kernel of the foreground using an analytical covariance function characterized by a set of hyperparameters, such as the variance and coherence length scale. However, the \HI\ 21\,cm signal from the EoR is governed by a complex set of astrophysical and cosmological parameters, resulting in intricate spatial and spectral fluctuations. Thus, the analytic covariance function relies on fixed functional forms, which are often insufficient for capturing the rich morphological complexity of the intrinsic cosmological signal \cite{Kern2021}. To address this limitation, we adopted the method introduced by Mertens et al. \cite{Mertens2024}, which replaces the standard analytic kernel with a variational auto-encoder kernel trained on simulations of the 21\,cm signal. This learned kernel provides a physically motivated prior for describing the signal of interest.

However, estimating the signal components $\mathbf f_i$ requires first optimizing the hyperparameters of the covariance functions to ensure the best model fit. The optimization is achieved by maximizing the log-marginal likelihood\,(LML), which is the likelihood function marginalized over all latent functions and is expressed as:
\begin{equation}
    \log P(\mathbf{d}|\nu,\theta) = -\frac{1}{2} \mathbf{d}^{\mathrm{T}}\mathbf{K}^{-1}\mathbf{d} - \frac{1}{2}\log|\mathbf{K}| - \frac{N}{2}\log2\pi
    \label{equ:LML}
\end{equation}
Here, $\theta$ denotes the set of model hyperparameters, and $N$ is the number of data points. Each term in Equation~\ref{equ:LML} has distinct physical and statistical interpretations. The first term quantifies the goodness-of-fit by penalizing the residuals between the model and observations. The second term serves as a regularization penalty that reflects model complexity. The third term ensures proper normalization of the probability density function.
\begin{figure}[!htbp]
    \centering
    \includegraphics[width=\linewidth]{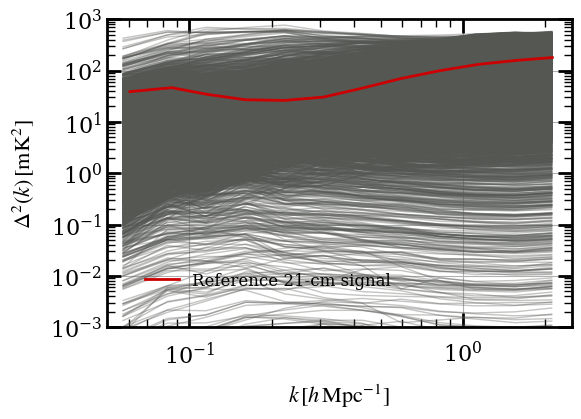}
    \caption{The spherically averaged power spectra for a selected subset of 500 simulations drawn from the total 5000 \textsc{21cmFAST} training ensemble. The 21\,cm reference signal is highlighted in red.}
    \label{fig:vae_ps_training}
\end{figure}
\subsection{ML-trained 21\,cm kernel}\label{sec:vae_training}
To construct a low-dimensional representation of the 21\,cm signal covariance, we require a large ensemble of mock 1D power spectra to train the VAE kernel. We utilized the semi-numerical code \textsc{21cmFAST} \cite{Murray2020} to generate a training dataset comprising $5,000$ distinct astrophysical models of EoR 21\,cm light-cones at our target redshift ($z \sim 9$). We adopt the parameterization introduced by Park et al. \cite{Park2019MNRAS.484..933P} with the following range of astrophysical parameters. These parameters are $\log{f_\star},~\alpha_\star,~\log{f_{\rm esc,10}},~\alpha_{\rm esc,10},~M_{\rm turn},$ and  $ t_{\star}$. Because each parameter set defines a distinct reionization scenario, enabling the VAE to capture a wide range of 21‑cm morphologies. We generated a parameter set of $5000$ Latin hypercube samples and simulated 21\,cm brightness temperature light cones. We computed the 1D power spectra from these boxes on the fly and saved them to prepare for subsequent training. The 1D power spectra were normalized before being converted into a covariance training dataset. This normalization ensures that the VAE learns only the shape of the 21\,cm power spectrum, allowing us to treat its variance as a separate free parameter. Our VAE training procedure follows the framework described by Mertens et al. \cite{Mertens2024}, which provides a comprehensive mathematical formulation of the VAE architecture in the context of 21‑cm cosmology. The VAE hyperparameters used in our training are summarized in Table~\ref{tab:vae model_trainning}. Figure~\ref{fig:vae_ps_training} illustrates the spherically averaged power spectra for a subset of the 5,000 \textsc{21cmFAST} simulations used in the training dataset.

\begin{table}
\centering
\renewcommand{\arraystretch}{1.1}
\caption{Details of the VAE network architecture used in this study. The VAE model was implemented using the \textsc{ps\_eor} package, which was built on the \textsc{PyTorch} framework. The model was trained on the 1D power spectra of simulated 21\,cm signals to learn and effectively encode the underlying covariance structures.} \label{tab:vae model_trainning}
\begin{tabular}{cc}
\toprule
Encoder hidden dimension & 20,20,20 \\
Latent dimension & 2 \\
Decoder hidden dimension & 20,20,20 \\
Batch size & 128 \\
Epoch & 20000 \\
Regularization constant & 1e$^{-4}$ \\
Learning rate & 5e$^{-3}$ \\
Optimizer & Adagrad \\
\bottomrule
\end{tabular}
\end{table}
\subsection{Candidate GPR models}\label{sec:gpr_models}
The GPR method is well established for modelling and subtracting residual foregrounds \cite{Mertens2020, Mertens2024, Mertens2025} in 21\,cm observations from various radio interferometric experiments. However, comparing models across studies is challenging because GPR frameworks are typically customized with parameters specific to each dataset. Furthermore, GPR is non-parametric in nature, even a small change in a single hyperparameter can lead to unpredictable shifts in the underlying latent functions. These factors make it challenging to identify the optimal configuration of the target sky patch. Building upon the parameterized models explored by Liu et al. \cite{Liu2026_MLGPR}, we extend their comparative analysis to realistic SKA1-Low $AA^{*}$ mock observations. By explicitly incorporating instrumental and environmental systematics into this simulation enables us to assess the robustness of residual foreground mitigation under the realistic observational conditions described in Section~\ref{sec:syn}.

In this work, we consider four distinct Bayesian GPR models -- \manalytic{}, \ubase{}, \mwedge{}, and \anoise{} -- to effectively reconstruct the total-sky intensity in SKA observations. These models share identical treatments of thermal noise and intrinsic foregrounds but differ in their parametrization for the characterization of the mode-mixing foreground components and the 21\,cm signal. Next, we describe the covariance kernels used for each component of the ML-GPR framework.

Intrinsic foregrounds: This component represents large-scale spectrally smooth astrophysical emission from extragalactic sources within the field of view. Across all frameworks, we model this component using a Median Radial Basis Function\,(MRBF) covariance kernel, obtained by setting $\eta = \infty$ in Equation~\ref{eq:matern}. To capture its spectral smoothness, we assign the MRBF kernel a large length scale of 100\,MHz, which is much larger than the total simulation bandwidth. 

Mode-mixing foregrounds: The mode-mixing term captures chromatic spectral structure introduced by the instrument response, which produces shorter frequency-coherence scales and manifests as the wedge in the 2D power spectrum. We also describe this component with an MRBF kernel, whose coherence scale varies with baseline length, thereby accounting for the baseline dependence of foreground mode-mixing. We parameterize the mode-mixed coherence scale ($l_{\rm mix}(u)$) as
\begin{equation}
    l_{\rm mix}(u) =  \frac{f_m}{u~sin\theta_{\rm mix} + f_mC_{mix}}
    \label{equ:l_mix}
\end{equation}
where $u$ is the baseline length, $f_m$ is the central frequency of the observation. $\theta_\mathrm{mix}$ and $C_\mathrm{mix}$ control the mode-mixing scale and can be interpreted as the wedge angle and the wedge buffer, respectively \cite{Mertens2025}.

21\,cm signal: To model the 21\,cm signal covariance function, we employ a machine-learning-trained VAE kernel. This robust model is characterized by three parameters: two latent space dimensions ($x_1, x_2$) and a variance scaling factor ($\sigma_{21}^2$). Integrating this VAE-derived covariance into our GPR framework significantly enhances our ability to separate the 21\,cm signal from foreground contaminants and systematic errors. A detailed methodology of the VAE approach is available in Mertens et al. \cite{Mertens2024}, and the training of the VAE using \textsc{21cmFAST} simulations is discussed in Section~\ref{sec:vae_training}.

Thermal noise: This covariance is modelled as a diagonal matrix, with its variance estimated from the time-differenced Stokes $I$ visibility cube.

The four candidate GPR models combine these kernels in the following configurations:
\manalytic{}: In this model, we used analytical covariance functions to parameterize the mode-mixed foreground and 21\,cm signal components. For the 21\,cm signal covariance function, we choose an exponential kernel (EXP), corresponding to a Mat\'ern kernel with $\eta = 1/2$. For modelling the mode-mixing components, consider the baseline-dependent coherence scale as described in Equation~\ref{equ:l_mix}.

\ubase{}: In this model, the 21\,cm signal is represented with a VAE kernel trained on simulated 21\,cm signal data, replacing the analytical covariance of \manalytic{} with a physically motivated prior. However, it simplifies the mode-mixed foreground component by modelling it with a baseline-independent (uniform) variance and length scale.

\mwedge{}: We extend the \ubase{} model by incorporating the baseline-dependent mode-mixed coherence scale as described in Equation~\ref{equ:l_mix}. 

\anoise{}: In this model, we extend the \mwedge{} model by introducing a scaling factor, $\alpha_n$, to scale the thermal noise covariance. This additional free parameter compensates for any frequency-uncorrelated noise present in the data beyond the initial noise variance estimate.

A comprehensive summary of the various components of the GPR frameworks, along with their respective parameter, prior and posterior constraints, is listed in Table~\ref{tab:gpr_models}. All models are implemented using the \texttt{ps\_eor}\footnote{\url{https://gitlab.com/flomertens/ps_eor}} Python package.
\begin{table}[htbp]
\renewcommand{\arraystretch}{1.25}
\centering
\caption{Summary of the covariance functions for the different candidate GPR models, including their parameters, prior distributions, and posterior constraints. The length-scale parameters in the \mwedge{}, \anoise{}, and \manalytic{} models follow the wedge parameterization (Equation~\ref{equ:l_mix}). The 21\,cm signal is modeled using either a variational autoencoder (``VAE'') or an exponential kernel. The \anoise{} model additionally introduces a scaling of the noise covariance of the form $\alpha_{\mathrm{n}} \mathbf{K}_{\mathrm{n}}$ (see Section~\ref{sec:gpr_models}). The symbols $\mathcal{U}$ and $\mathcal{LU}$ denotes uniform and log$_{10}$-uniform priors, respectively. The posterior values represent the median along with the $68\%$ credible intervals obtained from nested sampling. }
\label{tab:gpr_models}
\begin{tabular}{|c|c|c|c|p{2.0cm}|}
\hline
\textbf{Model} & \textbf{Kernel Type} & \textbf{Parameter} & \textbf{Prior} & \multicolumn{1}{c|}{\textbf{Posterior}} \\
\hline\hline

\multirow{7}{*} \mwedge{} 
& $K_{\rm int}$ (RBF)
& $\sigma^{2}_{\rm int}$ 
& $\mathcal{LU}(-0.6, 1.6)$ 
& $-0.093^{+0.025}_{-0.025}$ \\
\cline{2-5}
& \multirow{3}{*}{$K_{\rm mix}$ (RBF)}
& $\sigma^{2}_{\rm mix}$ 
& $\mathcal{LU}(-2, -0.2)$ 
& $0.322^{+0.014}_{-0.013}$ \\
& & $\theta_{\rm mix}$ [rad] 
& $\mathcal{U}(0, 0.2)$ 
& $0.141^{+0.001}_{-0.001}$ \\
& & $C_{\rm mix}$ [$\mu~s$] 
& $\mathcal{U}(0, 0.1)$ 
& $0.034^{+0.002}_{-0.002}$ \\
\cline{2-5}
& \multirow{3}{*}{$K_{21}$ (VAE)}
& $\sigma_{\rm 21}^2$ 
& $\mathcal{LU}(-7, -2)$ 
& $-4.149^{+0.102}_{-0.083}$ \\
& & $x_1$ 
& $\mathcal{U}(-8, 8)$ 
& $-2.029^{+0.190}_{-0.258}$ \\
& & $x_2$ 
& $\mathcal{U}(-8, 8)$ 
& $3.368^{+0.238}_{-0.272}$ \\
\hline\hline

\multirow{10}{*} \anoise{}
& $K_{\rm int}$ (RBF)
& $\sigma^{2}_{\rm int}$ 
& $\mathcal{LU}(-0.6, 1.6)$ 
& $-0.092^{+0.026}_{-0.024}$ \\
\cline{2-5}
& \multirow{3}{*}{$K_{\rm mix}$ (RBF)}
& $\sigma^{2}_{\rm mix}$ 
& $\mathcal{LU}(-2, -0.2)$ 
& $0.322^{+0.013}_{-0.014}$ \\
& & $\theta_{\rm mix}$ [rad] 
& $\mathcal{U}(0, 0.2)$ 
& $0.141^{+0.001}_{-0.001}$ \\
& & $C_{\rm mix}$[$\mu~s$] 
& $\mathcal{U}(0, 0.1)$ 
& $0.034^{+0.002}_{-0.002}$ \\
\cline{2-5}
& \multirow{3}{*}{$K_{21}$ (VAE)}
& $\sigma_{\rm 21}^2$ 
& $\mathcal{LU}(-7, -2)$ 
& $-4.139^{+0.096}_{-0.086}$ \\
& & $x_1$ 
& $\mathcal{U}(-8, 8)$ 
& $-1.975^{+0.202}_{-0.246}$ \\
& & $x_2$ 
& $\mathcal{U}(-8, 8)$ 
& $-3.734^{+0.239}_{-0.279}$ \\
\cline{2-5}
& $K_{n}$ (White noise) 
& $\alpha_n$ 
& $\mathcal{U}(0.5, 1.5)$ 
& $0.964^{+0.007}_{-0.007}$ \\
\hline\hline

\multirow{8}{*} \ubase{}
& $K_{\rm int}$ (RBF)
& $\sigma^{2}_{\rm int}$ 
& $\mathcal{LU}(-0.6, 1.6)$ 
& $-0.060^{+0.014}_{-0.012}$ \\
\cline{2-5}
& \multirow{2}{*}{$K_{\rm mix}$ (RBF)} 
& $\sigma^{2}_{\rm mix}$ 
& $\mathcal{LU}(-2, -0.2)$ 
& $-0.528^{+0.007}_{-0.007}$ \\
& & $\ell_{\rm mix}$ [MHz] 
& $\mathcal{U}(2, 10)$ 
& $2.429^{+0.006}_{-0.006}$ \\
\cline{2-5}
& \multirow{3}{*}{$K_{21}$ (VAE)} 
& $\sigma_{\rm 21}^2$ 
& $\mathcal{LU}(-7, -2)$ 
& $-5.795^{+0.083}_{-0.027}$ \\
& & $x_1$ 
& $\mathcal{U}(-8, 8)$ 
& $7.110^{+0.630}_{-2.111}$ \\
& & $x_2$ 
& $\mathcal{U}(-8, 8)$ 
& $6.987^{+0.653}_{-1.555}$ \\
\hline\hline

\multirow{6}{*} \manalytic{}
& $K_{\rm int}$ (RBF)
& $\sigma^{2}_{\rm int}$ 
& $\mathcal{LU}(-0.6, 1.6)$ 
& $-0.075^{+0.022}_{-0.021}$ \\
\cline{2-5}
& \multirow{3}{*}{$K_{\rm mix}$ (RBF)} 
& $\sigma^{2}_{\rm mix}$ 
& $\mathcal{LU}(-2, -0.2)$ 
& $-0.323^{+0.011}_{-0.012}$ \\
& & $\theta_{\rm mix}$ [rad] 
& $\mathcal{U}(0, 0.2)$ 
& $0.135^{+0.001}_{-0.001}$ \\
& & $C_{\rm mix}$ [$\mu~s$] 
& $\mathcal{U}(0, 0.1)$ 
& $0.048^{+0.002}_{-0.002}$ \\
\cline{2-5}
& \multirow{2}{*}{$K_{21}$ (EXP)} 
& $\sigma^{2}_{\rm 21}$ 
& $\mathcal{LU}(-7, -2)$ 
& $-5.560^{+0.012}_{-0.019}$ \\
& & $\ell_{\rm 21}$ [MHz] 
& $\mathcal{U}(0.1, 0.4)$ 
& $-0.397^{+0.002}_{-0.014}$ \\
\hline
\end{tabular}
\end{table}

Each sky component is modelled with an associated set of hyperparameters, for which prior ranges were assigned and iteratively refined throughout the model development process. The ML-GPR framework is applied to the source-subtracted gridded visibility data described in Section~\ref{compact_source_sub} to mitigate residual foreground contamination further and extract the underlying cosmological 21\,cm signal. We quantified parameter uncertainty via the posterior distributions of the kernel hyperparameter,  which are then propagated through subsequent analyses to assess their impact on power-spectrum measurements. For each GPR model, the Bayesian evidence and posterior probability distributions of the kernel hyperparameters are estimated using the nested sampling algorithm \textsc{UltraNest}. The resulting posterior distributions are presented in Appendix~\ref{appendix:posterior_density_distribution}. The priors and posterior values represent our most promising understanding of the spectral behaviours of the foregrounds in the observed sky. This framework ensures the effective extraction of the cosmological 21\,cm signal from astrophysical foregrounds and residual systematic contamination in radio observations. To validate the robustness of both parameter inference and signal recovery, we perform an injection test. In this process, mock 21\,cm signals are injected into the data and verified for accurate recovery using our component separation framework to evaluate the fidelity of the reconstruction (see the detailed procedure in Section~\ref{sec:signal_injection_test}).
\section{Results}\label{sec:results}
In this section, we assess the performance of different ML-GPR models to recover the cosmological 21\,cm signal from realistic SKA1-Low simulations. We demonstrate how these models can reconstruct the EoR power spectrum at $z = 9$ in a robust and relatively unbiased manner from pre-processed SKA data.

\subsection{Bayesian evidence}
\begin{figure}
    \centering
    \includegraphics[width=0.8\linewidth]{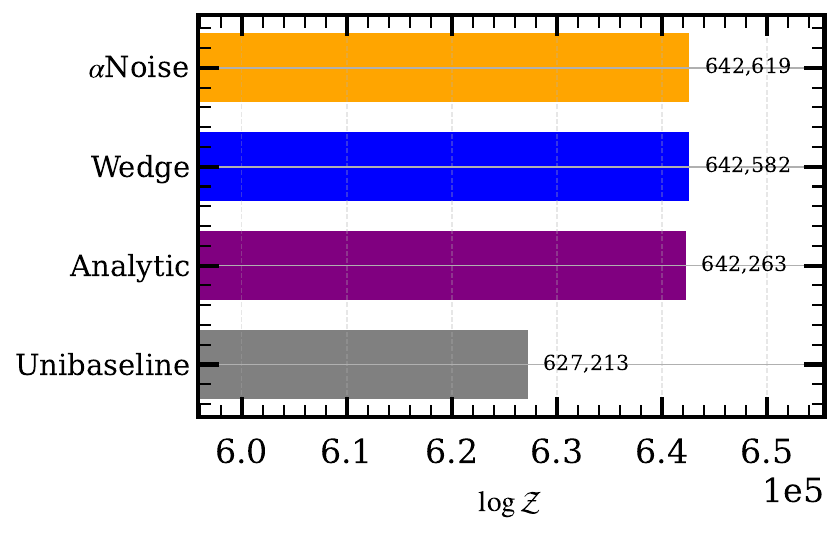}
    \caption{Comparison of Bayesian log-evidence ($\log\mathcal{Z}$) for the four candidate GPR models. A higher value indicates a statistically preferred model that better describes the observed data, providing a better balance between goodness-of-fit and model complexity. The \anoise{} model achieves the highest evidence, followed by \mwedge{}, \manalytic{}, and \ubase{}.}
    \label{fig:gpr_evidence}
\end{figure}

We assessed the relative performance of the four models (\manalytic{}, \ubase{}, \mwedge{}, and \anoise{}) using Bayesian evidence\,($\log\mathcal{Z}$). Bayesian evidence provides a global measure of model fit, quantifying how well each model explains the observed data, while inherently accounting for its complexity. In this framework, models that achieve a better balance between goodness-of-fit and parameter economy are statistically preferred. Figure~\ref{fig:gpr_evidence} shows the comparison of the Log Bayesian evidence obtained for four candidate GPR models. Among the candidates, the \anoise{} model yields the highest Bayesian evidence, providing the most statistically favoured description of the observed data. This model extends the \mwedge{} parameterization by introducing an additional noise-scaling component, providing a crucial degree of freedom to account for frequency-uncorrelated noise fluctuations. Thus, the combined kernel structure captures both the baseline-dependent foreground mode-mixing and the stochastic noise contribution more effectively than the \mwedge{} model. The relative difference in log evidence ($\Delta\log\mathcal{Z} = 36.72$) represents a decisive, statistically significant improvement in modelling the total-sky intensity. In addition, the log-evidence values for \manalytic{} and \ubase{} are lower than \anoise{} by $0.056\%$ and $2.4\%$, respectively. The relatively small difference between \manalytic{} and \anoise{} suggests that the analytic 21\,cm covariance prescription already captures a substantial fraction of the spectral structure present in the data.

However, the additional flexibility introduced in \anoise{} provides a more accurate representation of the residual stochastic components, thereby improving the overall fit. By contrast, the poor performance of the \ubase{} model demonstrates the importance of incorporating baseline-dependent foreground structure into the covariance model. The catastrophic reduction in Bayesian evidence indicates that neglecting baseline dependence in foreground mode-mixing renders a model incapable of describing the observed data. This result highlights that physically motivated covariance structures are essential for robust foreground mitigation in realistic SKA1-Low $AA^{*}$ observations.

\begin{figure}
    \centering
    \includegraphics[width=\linewidth]{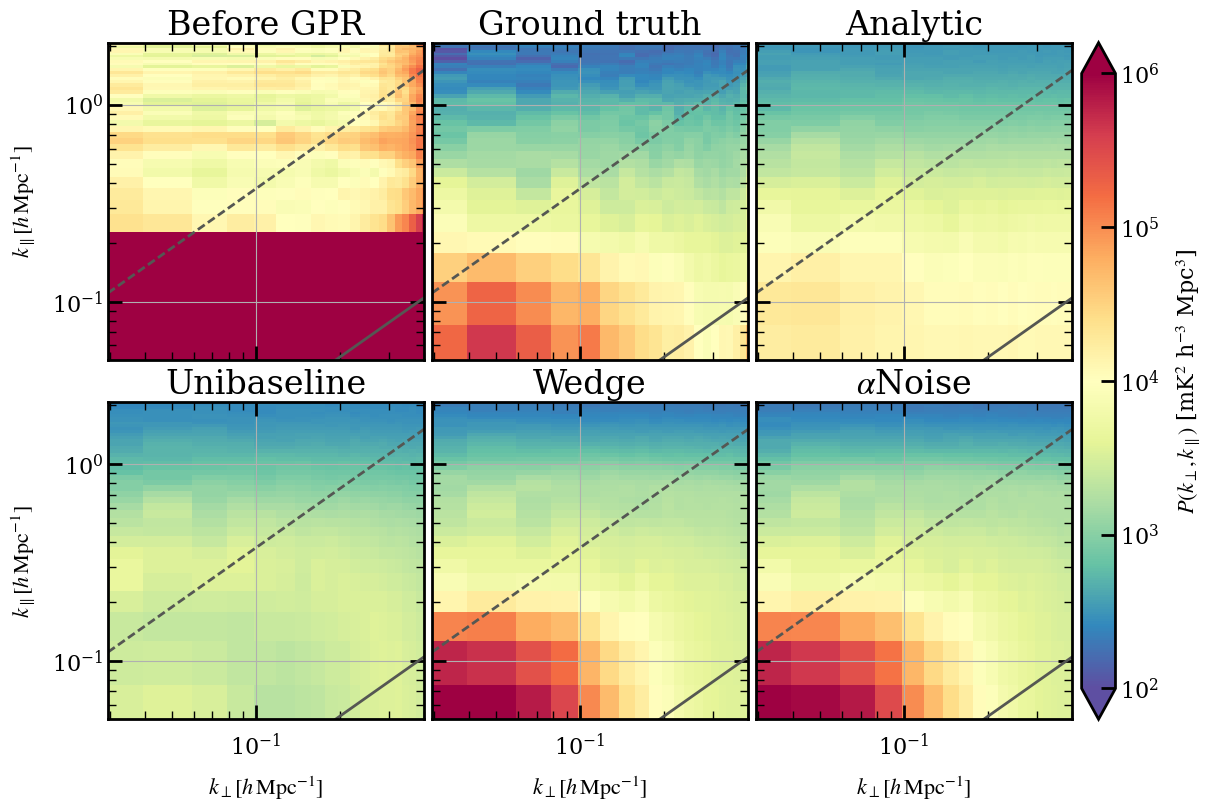}
    \caption{Recovered cylindrical average 2D power spectra from the source-subtracted SKA-Low AA$^{*}$ data, illustrating the efficacy of the four ML-GPR models in reconstructing the 21\,cm signal. The top-left panel shows the data before GPR, contaminated by foreground leakage into the EoR window, and the top‑middle panel shows the ground‑truth \HI\ signal 2D PS. The remaining panels present the reconstructed residual power spectra for the \ubase{}, \mwedge{}, \manalytic{}, and \anoise{} models, after correcting for the thermal-noise bias. Both the \mwedge{} and \anoise{} models most closely match the ground truth.}   
    \label{fig:2d_ps_gpr}
\end{figure}
\begin{figure}
    \centering
    \includegraphics[width=\linewidth]{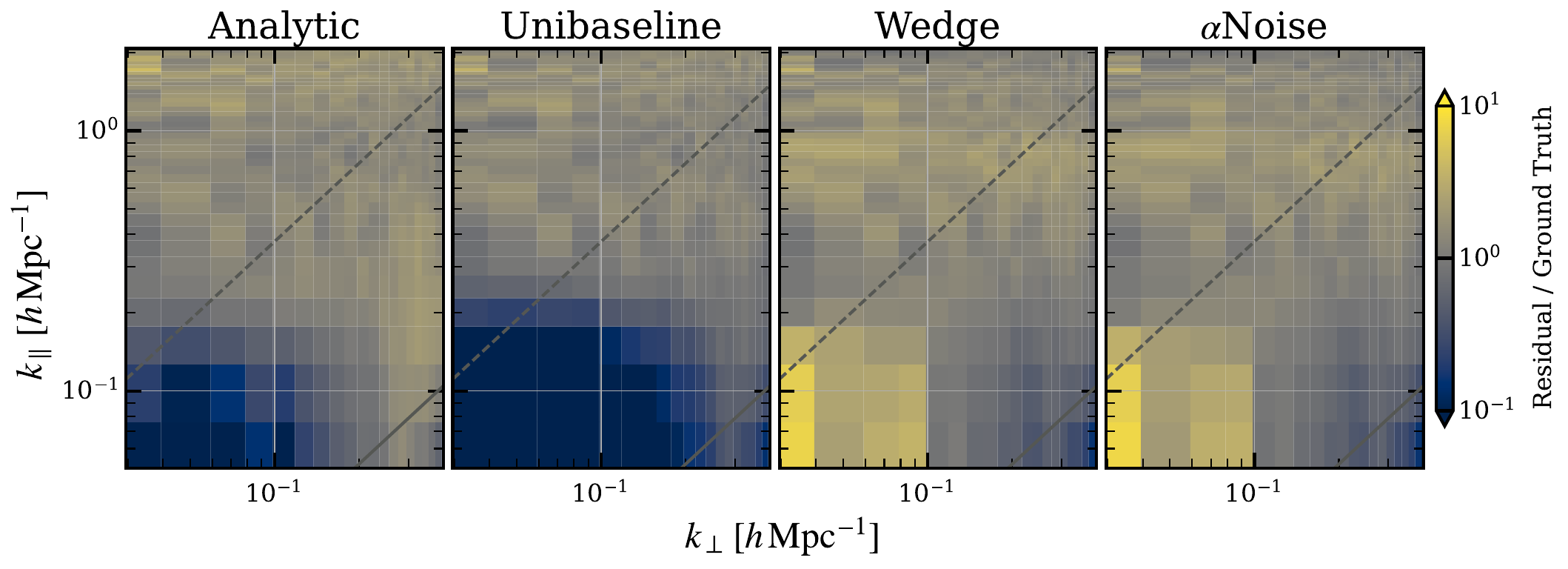}
    \caption{Comparison of the ratio of the recovered 21\,cm power spectrum to the ground truth for four GPR models.}
    \label{fig:ps2d_ratio}
\end{figure}

\subsection{Power spectrum analysis} \label{sec:ps_analysis}
We present the efficacy of the four GPR models in the cylindrically averaged 2D residual power spectrum estimated from the source-subtracted SKA-Low data, as illustrated in Figure~\ref{fig:2d_ps_gpr}. After estimating the posterior distributions of the kernel hyperparameters for each model via nested sampling, the residual foreground was subtracted from the pre-processed SKA-observed dataset. Finally, to extract the cosmological signal, a bias correction was applied by subtracting the estimated thermal noise power spectrum. The top-left panel of Figure~\ref{fig:2d_ps_gpr} depicts the power spectra prior to applying ML-GPR, showing significant contamination at higher $k_{\parallel}$ modes, caused by the combined effects of ionospheric phase errors and instrumental effects. These residual systematic errors cause foreground power to leak out of the wedge and into the EoR window, thereby potentially obscuring the underlying cosmological signal. For comparison, the top-middle panel of Figure~\ref{fig:2d_ps_gpr} shows the ground-truth \HI\ signal power spectrum observed by SKA-Low AA* configuration, free from foreground contamination, instrumental noise, and residual systematics. The remaining panels in figure~\ref{fig:2d_ps_gpr} illustrate the efficacy of the ML-GPR method in reconstructing residual power spectra. From the visual inspection, we can see that the ML-GPR method significantly improved the recovery of the 21\,cm signal. However, the lower $k_{\parallel}$ modes are still dominated by the residual foreground power spectrum.

To quantify reconstruction fidelity across scales, Figure~\ref{fig:ps2d_ratio} shows the ratio of the recovered to ground-truth 21\,cm power spectra for each of the four GPR models, thereby highlighting their scale-dependent foreground-separation performance. Although \manalytic{} model includes a physically motivated baseline-dependent mode-mixing coherence scale, which captures the $k_{\perp}$-dependent variation in foreground power. However, the exponential covariance function used for the 21\,cm signal is inadequate for capturing the shape of the 21\,cm signal frequency-frequency covariance \cite{Kern2021}. This causes the 21\,cm kernel to become degenerate with the foreground component, leading the GPR framework to absorb part of the cosmological signal into the foreground model and leading to substantial signal loss. 

While \ubase{} model, which replaces the exponential kernel with the VAE-trained covariance prior, substantially improves the modelling of the intrinsic $k_{\perp}$-dependent structure of the 21\,cm signal. Despite this improvement, its mode-mixing component lacks explicit baseline dependence, preventing it from adequately capturing foreground power variations across the $k_{\perp}$ direction. As a result, this model overestimates the foreground wedge at large angular scales and underestimates it at small angular scales. This mismatch overestimates the foreground wedge on large angular scales and underestimates it on small scales, forcing residual foreground power to leak into the 21\,cm component. This leakage manifests as the strong signal suppression at small $k_{\perp}$ visible in the second panel of Figure~\ref{fig:ps2d_ratio}. In contrast, the \mwedge{} and \anoise{} models successfully combine the VAE-based covariance kernel with baseline-dependent covariance priors. This combination results in improved foreground separation, significantly reduced bias, and more reliable signal recovery. While the current set of covariance priors effectively removes most of the systematic errors induced by gain calibration errors and ionospheric residuals, a small but systematic bias remains visible for both the \mwedge{} and \anoise{} models, particularly along $k_{\parallel}$ at low $k_{\perp}$. This emphasizes the importance of properly accounting for all components in the covariance prior model and the need to develop more physically motivated kernels to capture instrument-induced mode mixing and ionospheric effects. Both models exhibit comparatively small reconstruction uncertainties. As illustrated in Figures~\ref{fig:2d_ps_gpr} and \ref{fig:ps2d_ratio}, the remaining uncertainties are largest at low $k_{\parallel}$, where foreground contamination dominates, and at small angular scales, where the signal-to-noise ratio is lowest.

\begin{figure}[!ht]
    \centering
    \includegraphics[width=\linewidth]{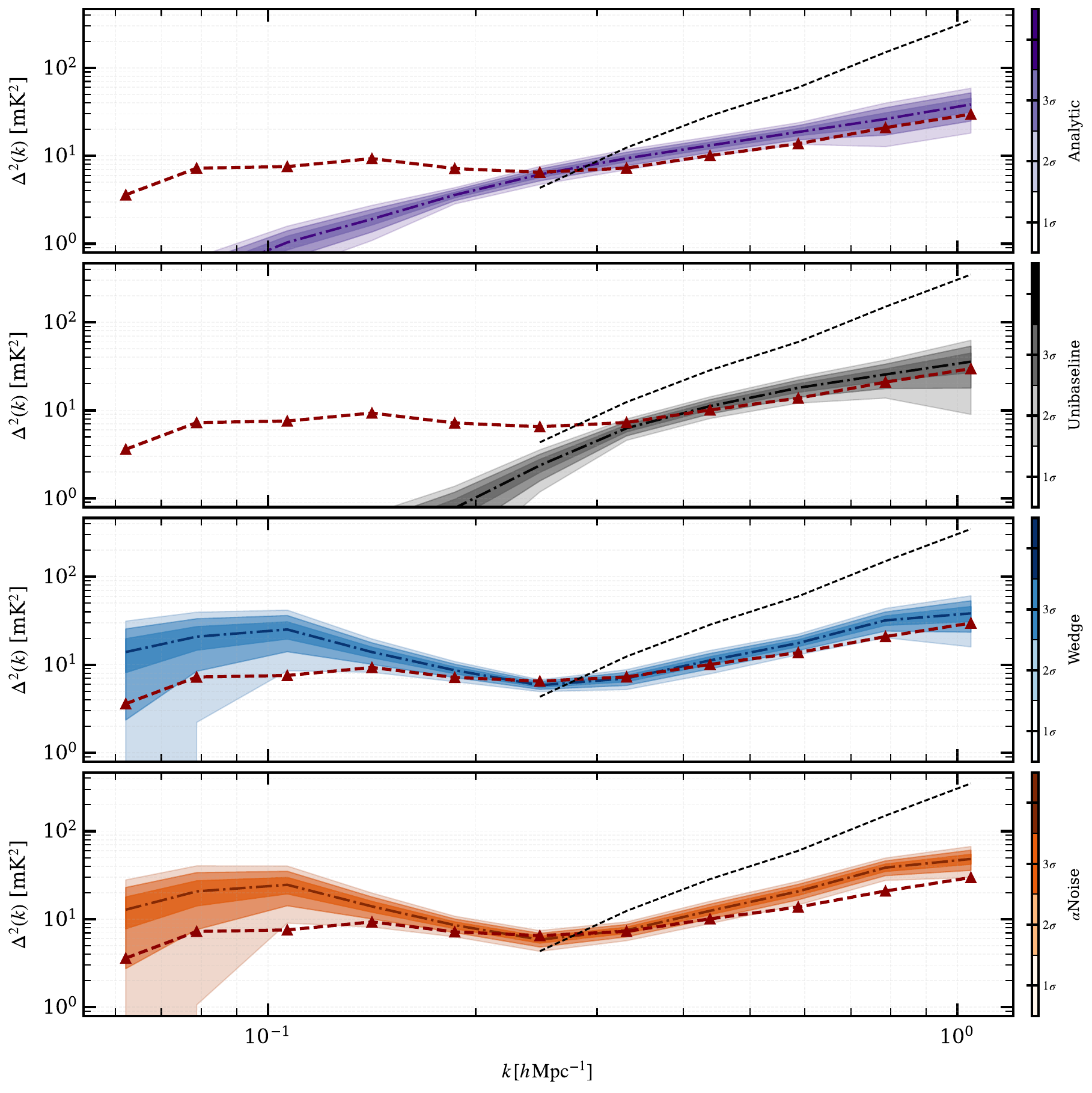}
    \caption{Recovered dimensionless spherically averaged 1D power spectra for the four candidate ML-GPR models. The four panels (top to bottom) present the 21\,cm signal power spectra, $\Delta^2_{\text{rec}}(k)$, recovered by the \manalytic{} (purple), \ubase{} (gray), \mwedge{} (blue), and \anoise{} (orange) GPR models. The ground-truth 21\,cm signal ($\Delta^2_{21}(k)$, red dash-triangles) and thermal noise limit ($\Delta^2_{\text{n}}(k)$, dashed line) are shown for comparison. For each model, the dash-dot line denotes the posterior mean, with shaded regions highlighting the $1\sigma$, $2\sigma$, and $3\sigma$ credible intervals.}
    \label{fig:1d_ps_gpr}
\end{figure}

Next, we compared the performance of four ML-GPR models in recovering the spherically averaged 1D power spectrum against the ground truth. Figure~\ref{fig:1d_ps_gpr} presents the recovered 1D power spectra with uncertainties estimated from the source-subtracted SKA-Low data across different ML-GPR decomposition models and compares them with the ground-truth power spectrum. The shaded regions indicate the $1\sigma$, $2\sigma$, and $3\sigma$ uncertainty intervals, which account for the spread of the hyperparameter posterior distribution. Overall, the \mwedge{} and \anoise{} models provide consistent, relatively unbiased reconstructions and achieve comparable performance, with the smallest residuals across all accessible $k$-modes for $0.06 \leq k \leq 1.0$~h\,Mpc$^{-1}$. In contrast, the \manalytic{} and \ubase{} models systematically over-subtract the foreground components, thereby poorly recovering the residual power spectrum on larger scales where $k < 0.25$~h\,Mpc$^{-1}$ as shown in figure~\ref{fig:1d_ps_gpr}. Although the recovery is not perfect, the reconstructed power for \mwedge{} and \anoise{} models remains consistent with the ground-truth signal within the $2\sigma$ credible interval for almost all the $k$-modes.

\begin{figure}
    \centering
    \includegraphics[width=\linewidth]{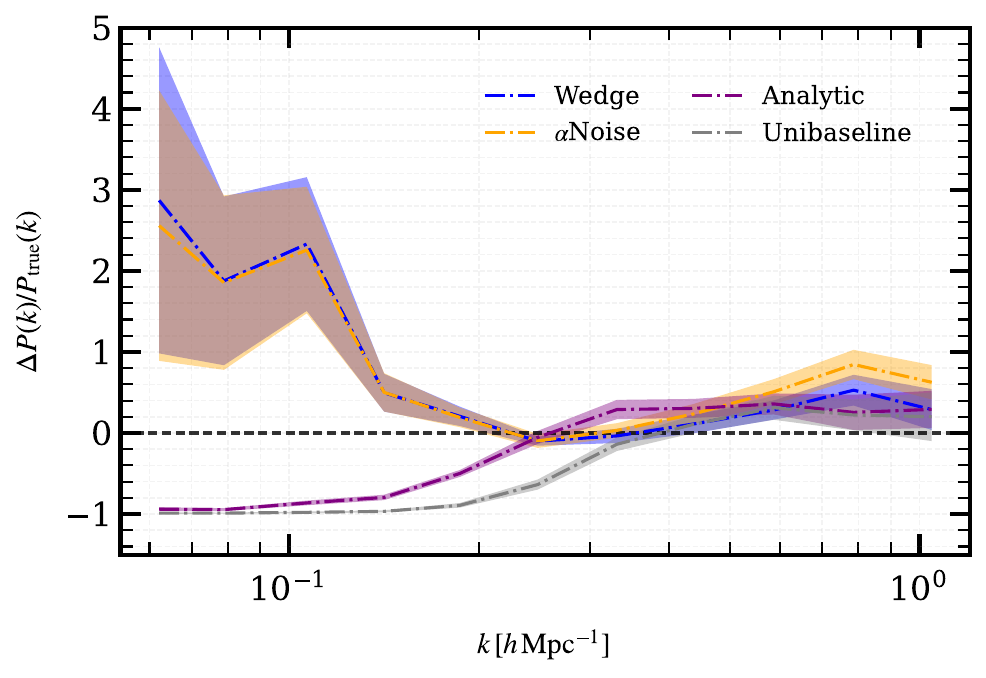}
    \caption{Comparison of the fractional residual power spectra, $\Delta P(k)/P_{\rm true}(k)$, as a function of sampled $k$-modes. The colour scheme matches that of Figure~\ref{fig:1d_ps_gpr}. The dashed horizontal line indicates zero residual, with positive (negative) values corresponding to an overestimation (underestimation) of the recovered 21\,cm signal relative to the ground truth. Shaded regions denote the corresponding $1\sigma$ uncertainty intervals.}
    \label{fig:1d_relative_ps}
\end{figure}

Among the four GPR models, the \anoise{} model achieves the most accurate signal recovery at $k = 0.33~h\,\mathrm{Mpc}^{-1}$. It predicts a residual power of $7.49\,\mathrm{mK}^{2}$, overestimating the $7.26\,\mathrm{mK}^{2}$ ground-truth signal by a mere $3.19\%$. The \mwedge{} model achieves a similarly competitive, albeit slightly conservative, reconstruction with a fractional difference of $-3.60\%$. In contrast, the \ubase{} and \manalytic{} models exhibit substantially larger fractional differences of$-13.89\%$ and $28.95\%$, respectively, at the same k-mode, reflecting their systematically poorer recovery of the cosmological signal. These results demonstrate that the \anoise{} and \mwedge{} models provide the most reliable and relatively unbiased reconstructions of the cosmological 21\,cm signal, whereas the remaining two models exhibit noticeably poorer recovery performance. The fractional residual power spectrum for each model is shown in  Figure~\ref{fig:1d_relative_ps}. 

\subsection{Signal injection tests}\label{sec:signal_injection_test}
In this section, we perform a signal-injection test to assess the robustness of the four GPR models in recovering unbiased estimates of the 21\,cm power spectrum. Although the simulated dataset analysed in Section~\ref{sec:ps_analysis} already includes a cosmological 21\,cm signal, it represents only a single reionisation scenario. It is essential to test the robustness of the ML-GPR framework to determine whether it remains reliable against a diverse parameter space of plausible signal shapes and amplitudes. We generate synthetic 21\,cm signals and inject them into the pre-processed SKA1-Low dataset before the ML-GPR component-separation step. To avoid bias towards any particular signal morphology, we sample 25 distinct 21\,cm signal shapes. For each shape, we rescaled the signal power to $0.25$, $0.5$, $1.0$, and $2.0$ times the thermal noise power, yielding 100 independent realisations of the injected signal spanning a broad range of amplitudes. For each realisation, we synthesised a visibility cube that matched the target power spectrum, added it to the data, and processed it using the same priors and hyperparameter settings described in Section~\ref{sec:gpr_models}. To isolate the injected signal, we run ML-GPR separately on the uninjected data to obtain the residual power spectrum, then subtract it from the power spectrum recovered with injection. Comparing the result to the known injected spectrum for each signal shape quantifies the attenuation or bias introduced by the ML-GPR framework. To measure the fidelity of the recovery, we compute a $z$-score for each sampled $k$-mode, measuring the deviation of the recovered spectrum $\Delta^{2}_{\text{rec}}(k)$ from the known injected spectrum $\Delta^{2}_{\text{inj}}(k)$ in units of statistical uncertainty:
\begin{equation} 
    z_j(k) = \frac{\Delta^2_{\mathrm{rec}, j}(k) - \Delta^2_{\mathrm{inj}, j}(k)}{\sigma_{\mathrm{rec}, j}(k)}, \end{equation}
where $\Delta^2_{\mathrm{rec}, j}(k)$ is the isolated, noise-bias-corrected residual power spectrum recovered for the $j$-th injection, and $\sigma_{\mathrm{rec}, j}(k)$ is the estimated $1\sigma$ predictive uncertainty in that bin. We summarize performance with the RMS $z$-score across the $N_{\mathrm{inj}} = 100$ realizations:
\begin{equation} 
    z_{\mathrm{RMS}}(k) = \sqrt{\frac{1}{N_{\mathrm{inj}}}\sum_{j=1}^{N_{\mathrm{inj}}} z^{2}_j(k)},
    \label{eq:averaged_z_score} 
\end{equation}

\begin{figure}[!htbp]
    \centering
    \includegraphics[width=\linewidth]{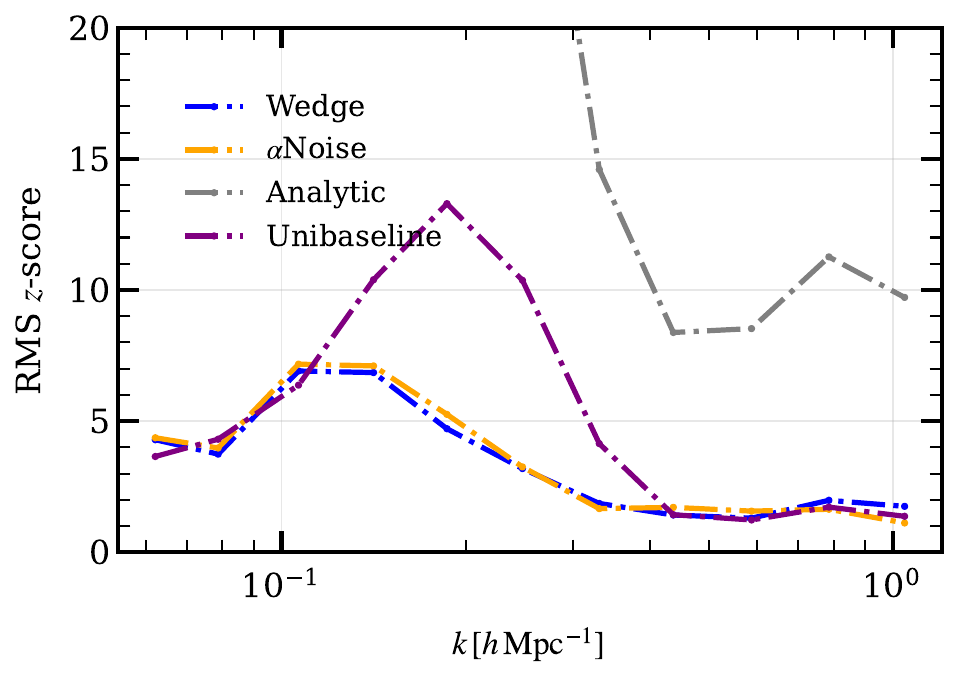}
    \caption{The RMS $z$-score as a function of the sampled $k$-modes. A lower RMS $z$-score indicates a more reliable and unbiased recovery of the 21\,cm signal at $k$-mode.}
    \label{fig:zscore_plot}
\end{figure}

Figure~\ref{fig:zscore_plot} shows that the RMS $z$-score across the sampled $k$-modes for the four models, demonstrating that the recovery accuracy is scale-dependent. The \mwedge{} and \anoise{} exhibit nearly identical performance and outperform both \manalytic{} and \ubase{}, indicating more accurate and less biased signal recovery. At large spatial scales ($k \lesssim 0.25~h\mathrm{Mpc}^{-1}$), however, all models exhibit elevated RMS $z$-scores, reflecting a persistent systematic bias. This degradation is particularly severe for injected signals characterised by flat power spectra or large-scale upturns. For these morphologies, the intrinsic spectral coherence of the 21\,cm signal approaches that of the mode-mixing foregrounds, which severely exacerbates the degeneracy between the cosmological component and residual systematic errors. Nevertheless, the posterior distributions provide tight, consistent constraints on the 21\,cm variance parameter, $\sigma_{21}^2$, across all models. This demonstrates that the ML-GPR framework can recover the cosmological signal with high fidelity across most accessible scales, reinforcing its promise as a robust foreground-mitigation approach for future SKA1-Low observations.

\section{Summary and conclusions}
\label{sec:summary}
We test the robustness of the ML-GPR framework for extracting the cosmological 21\,cm signal from simulated SKA1-Low AA* observations in the presence of realistic contaminants. To do this, we used the end-to-end \textsc{21cmE2E} simulation pipeline \citep{Mazumder2022, Pal2025a, Pal2025b} to generate mock data. The simulation incorporates a realistic sky model, beam responses, residual antenna-based gain calibration errors, ionospheric phase distortions, and leakage from bright out-of-field sources. The sky model comprises extragalactic point sources drawn from the GLEAM/LoBES survey, a 21\,cm signal generated with \texttt{21cmFAST}, and 1000\,h of observation equivalent instrumental noise. The instrument tracked the sky for 4\,hours with 10\,s integration times under quiet ionospheric conditions. For simplicity, we omit Galactic diffuse synchrotron emission and polarization leakage, and assume an identical primary beam response for all stations. From these simulated visibilities, we reconstructed high-resolution synthetic images using a uniform weighting scheme using \textsc{WSCLEAN}. We identified point sources with \textsc{PyBDSF}, generated model visibility for them, and subtracted this directly from the raw SKA-Low AA* data. We then created dirty image cubes using natural weighting from this pre-processed data. To mitigate residual foregrounds from confusion noise and unmodelled systematics, we applied the ML-GPR framework. Since EoR signal recovery at large cosmological scales is highly sensitive to how well residual foregrounds are removed, we compared four GPR parameterizations -- \manalytic{}, \ubase{}, \mwedge{}, and \anoise{} -- to forecast performance for future SKA1-Low observations. The key findings of this study are summarized as follows:

\begin{itemize}
    \item Comparing Bayesian evidence, the \anoise{} model performs best and provides the most statistically favoured description of the observed data. This is mainly because its noise-scaling parameter can flexibly capture frequency-uncorrelated noise. The \mwedge{} model also performed competitively. In contrast, the \ubase{} model was strongly disfavoured because it fails to capture the baseline-dependent structure caused by foreground mode-mixing. This demonstrates that baseline-dependent foreground modelling is essential for reliably recovering the 21\,cm signal in realistic SKA1-Low observations.
    
    \item The recovered power spectra show higher uncertainties in $k$-modes dominated by either residual foregrounds or thermal noise. Both the \mwedge{} and \anoise{} models provided largely unbiased reconstructions of the 1D power spectrum across almost all accessible $k$-modes ($0.06 \leq k \leq 1.0~h\,\mathrm{Mpc}^{-1}$). Although the recovery is not perfect, the reconstructed power remains consistent with the ground-truth signal within the 2--3 $\sigma$ credible interval on almost all $k$-modes. At $k = 0.33~h\,\mathrm{Mpc}^{-1}$, these models recover the cosmological power spectrum in a robust and relatively unbiased manner with fractional differences of only $3.19\%$ (\anoise{}) and $-3.60\%$ (\mwedge{}) relative to the ground truth. In contrast, the \manalytic{} and \ubase{} models systematically over-subtracted the foreground components, leading to a significant underestimation of the cosmological signal on large scales ($k < 0.33~h\,\mathrm{Mpc}^{-1}$), as illustrated in Figure~\ref{fig:1d_ps_gpr}. 
   
\end{itemize}
Liu et al. \citep{Liu2026_MLGPR} evaluated that the \anoise{} and \mwedge{} models perform best for signal separation, while the \manalytic{} and \ubase{} models fail Bayesian null tests by producing spurious detections. However, their analysis relied on idealized simulations that did not account for instrumental and propagation effects, which remain a primary challenge in 21\,cm cosmology. Our findings advance this by demonstrating that, without signal suppression, the underlying 21\,cm signal could be reliably recovered using an \anoise{} model even under realistic, low-level systematic-limited SKA1-Low AA$^{*}$ observations. These results also demonstrate the effectiveness of VAE-based GPR models in handling systematic contaminants. When the inherent baseline-dependent covariance prior is included for the mode-mixing component, the 21\,cm power spectrum is recovered successfully, albeit with higher uncertainty. However, real observational data will inevitably contain additional systematic effects not captured in our simulations, which can further complicate residual foreground subtraction. For instance, our framework assumes a complete and accurate sky model, whereas actual observations suffer from incomplete or inaccurate source catalogues. Furthermore, we did not incorporate bright A-TEAM sources, non-identical element beam patterns, mutual coupling, or the effects of missing and flagged frequency channels. We plan to test these methods against such realistic contaminants in future work. As Mertens et al. \citep{Mertens2024} demonstrated, if an excess systematic component is omitted from the prior model, its contribution to the overall variance of the data is inevitably absorbed by other components. Because the VAE-based 21\,cm kernel is highly flexible, it is particularly prone to absorbing this excess variance. Therefore, constructing accurate covariance prior models that account for all data systematics is of utmost importance. This emphasizes the importance of physically motivated covariance modelling for reliable foreground mitigation and future detections of the cosmological 21\,cm signal with SKA-Low.

\acknowledgments
SKP acknowledges financial support from the Department of Science and Technology, Government of India, through the INSPIRE Fellowship [IF200312]. The authors acknowledge the use of facilities procured through the funding via the Department of Science and Technology, Government of India sponsored DST-FIST grant no. SR/FST/PSII/2021/162(C) awarded to the DAASE, IIT Indore. The authors would like to thank the Square Kilometre Array (SKA) team members for promptly answering queries related to SKA station beam modelling.
\textbf{Software}: This work is heavily based on the Python programming language (\url{https://www.python.org/}). The packages used here are \texttt{astropy} (\url{https://www.astropy.org/}, \cite{Astropy2013A&A...558A..33A, Price-whelan2018AJ....156..123A} ), \texttt{numpy} (\url{https://numpy.org/}), \texttt{h5py} (\url{https://www.h5py.org/}), \texttt{matplotlib} (\url{https://matplotlib.org/}), \texttt{scipy} (\url{https://scipy.org/}).\\
\appendix
\section{Posterior density distribution}\label{appendix:posterior_density_distribution}
The posterior values represent the median along with the $68\%$ credible intervals obtained from nested sampling.
\begin{figure*}
    \centering
    \includegraphics[width=\linewidth]{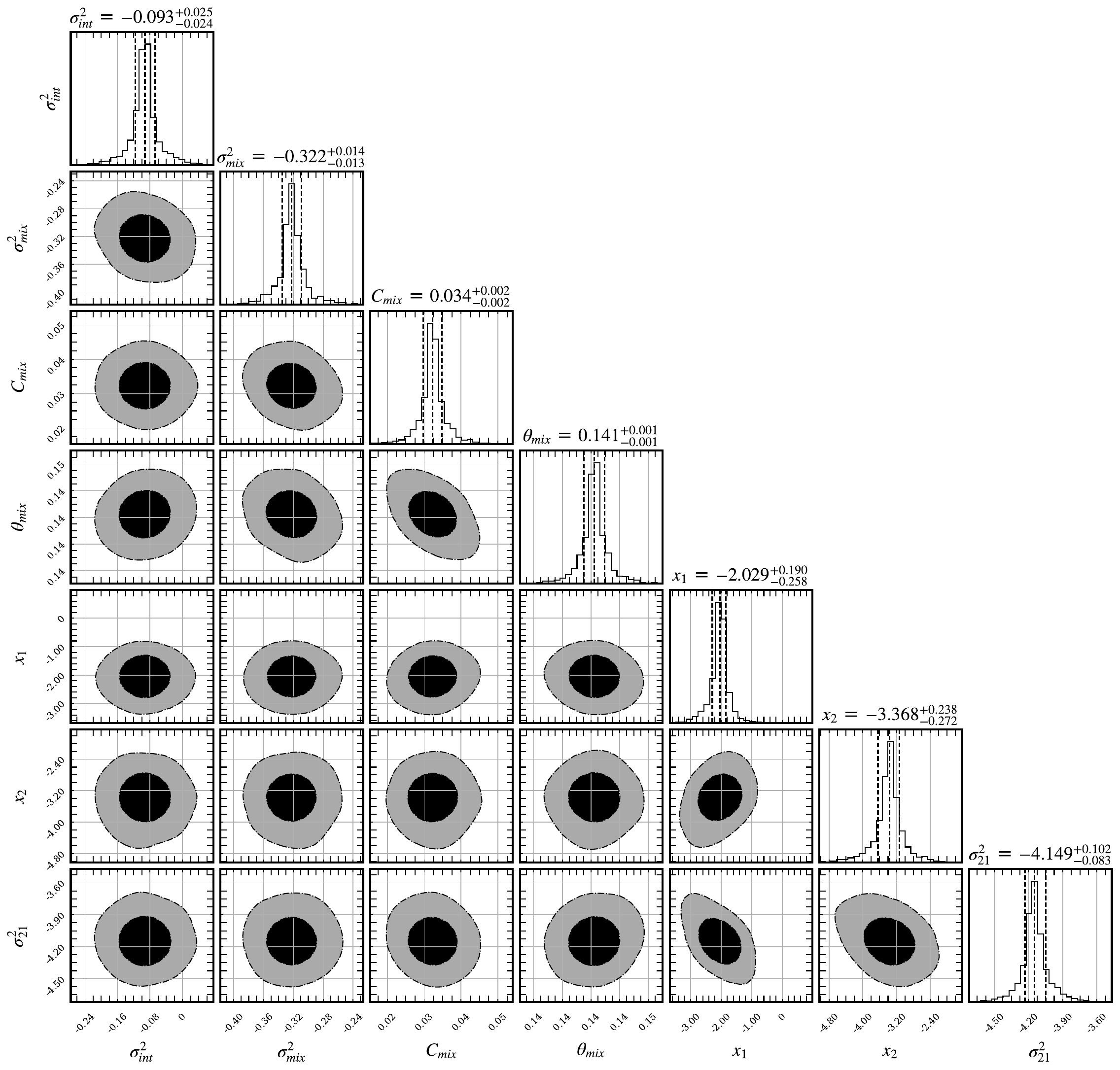}
    \caption{The posterior density distribution of the GPR hyperparameters after nested sampling for the \mwedge{} model.} \label{fig:wedge_posterior}
\end{figure*}

\begin{figure*}
    \centering
    \includegraphics[width=\linewidth]{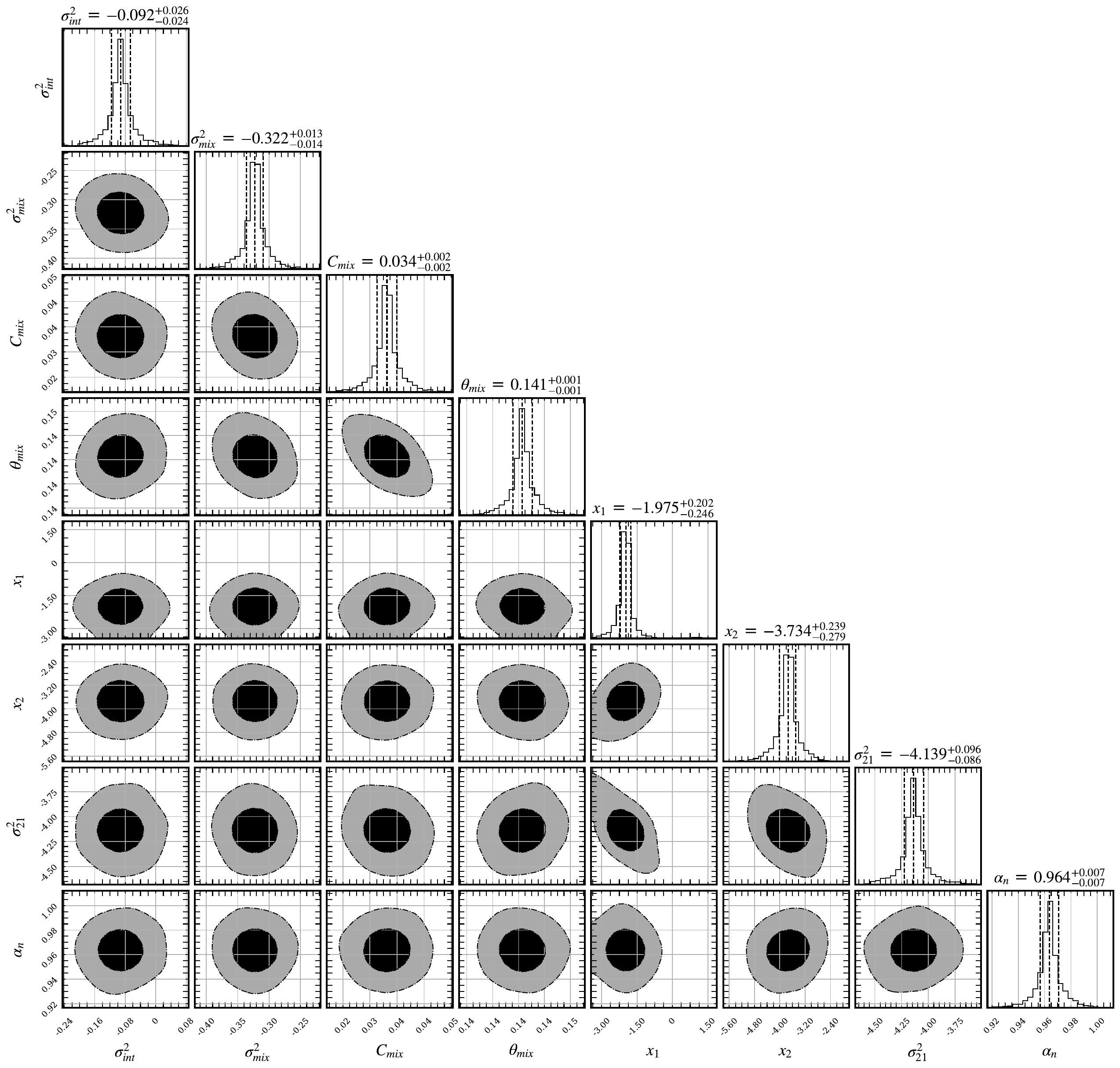}
    \caption{The posterior density distribution of the GPR hyperparameters after nested sampling for the \anoise{} model.} \label{fig:alpha_noise_posterior}
\end{figure*}

\begin{figure*}
    \centering
    \includegraphics[width=\linewidth]{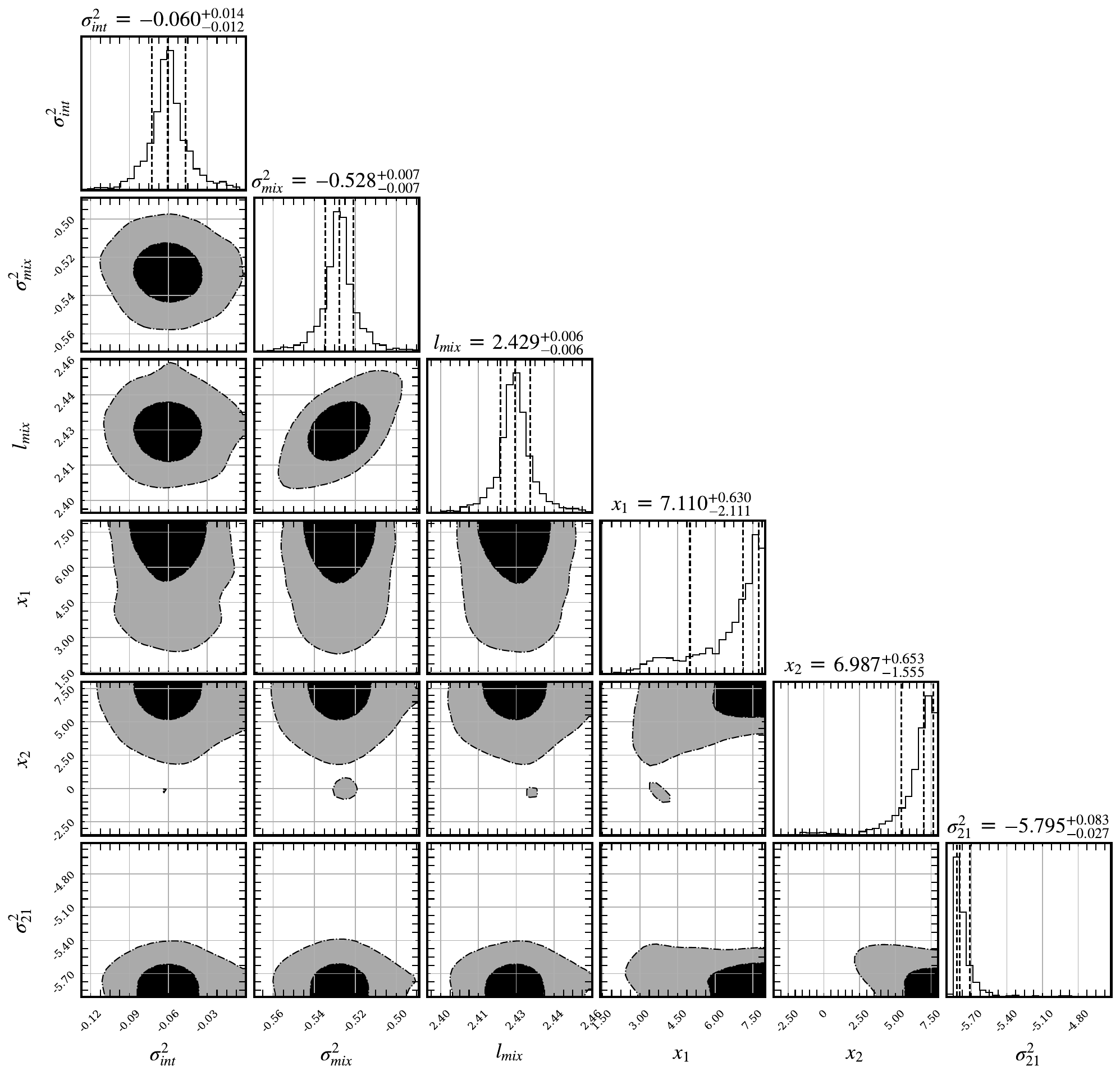}
    \caption{The posterior density distribution of the GPR hyperparameters after nested sampling for the \ubase{} model.} \label{fig:unibaseline_posterior}
\end{figure*}

\begin{figure*}
    \centering
    \includegraphics[width=\linewidth]{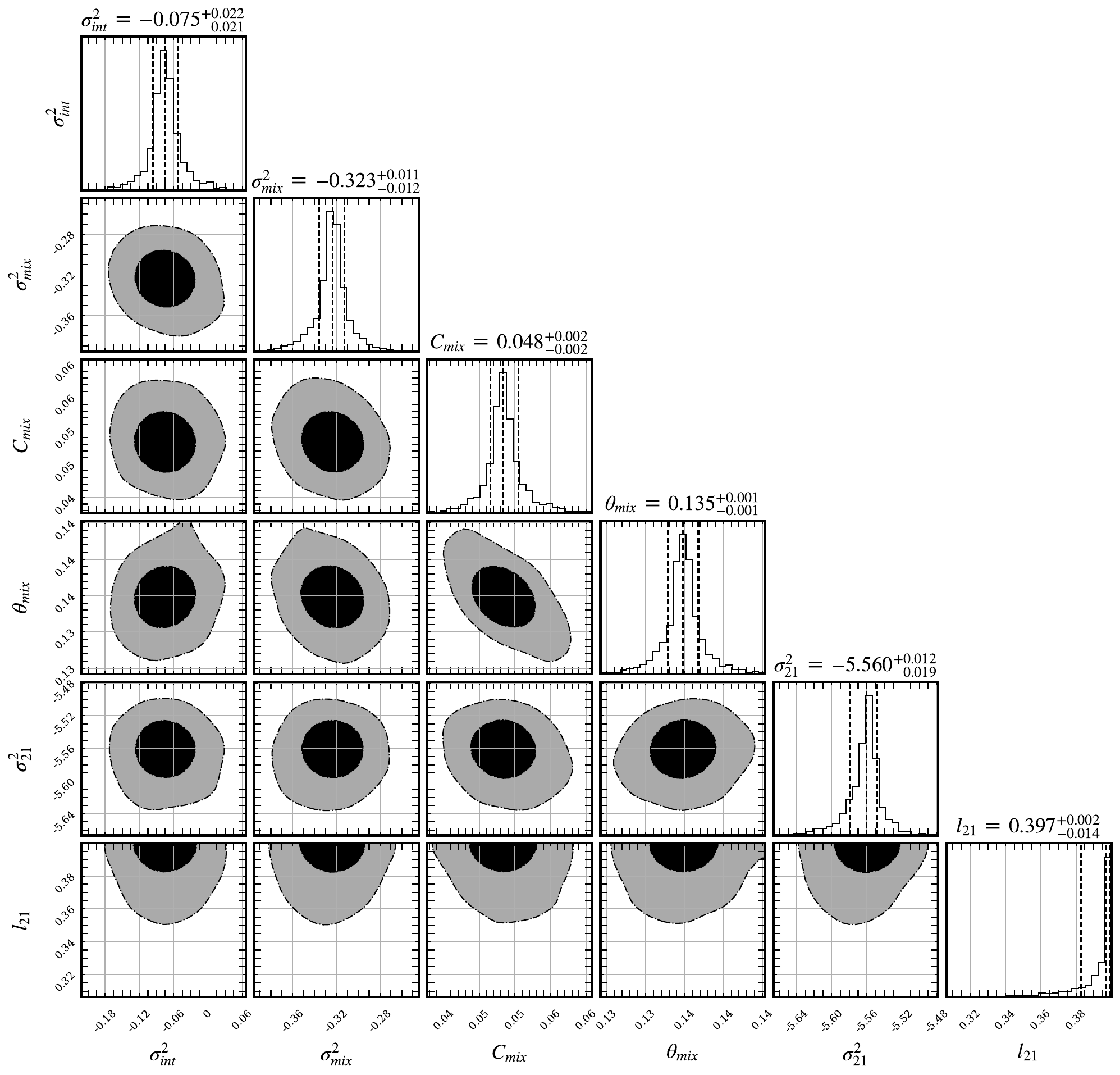}
    \caption{The posterior density distribution of the GPR hyperparameters after nested sampling for the \manalytic{} model.} \label{fig:analytic_posterior}
\end{figure*}

\bibliographystyle{JHEP}
\bibliography{main_ref}




\end{document}